\documentclass[aip,reprint]{revtex4-1}
\usepackage{graphicx}
\usepackage{txfonts}

\newcommand{\WHz}{\mathrm{\;W/\sqrt{\mathrm{Hz}}}}
\newcommand{\mum}{\mathrm{\;\mu m}}

\begin{document}

\title{Ultra-sensitive Super-THz Microwave Kinetic Inductance Detectors for future space telescopes} 

\author{J.J.A. Baselmans}
\email[]{J.Baselmans@sron.nl}
\affiliation{THz Sensing group, Microelectronics Department, Delft University of Technology, Mekelweg 1, 2628 CD - Delft, The Netherlands}
\affiliation{SRON - Netherlands Institute for Space Research, Niels Bohrweg 4 2333 CA - Leiden, The Netherlands}
\author{F. Facchin}
\affiliation{THz Sensing group, Microelectronics Department, Delft University of Technology, Mekelweg 1, 2628 CD - Delft, The Netherlands}
\author{A. Pascual Laguna}
\affiliation{SRON - Netherlands Institute for Space Research, Niels Bohrweg 4 2333 CA - Leiden, The Netherlands}
\author{J. Bueno}
\affiliation{SRON - Netherlands Institute for Space Research, Niels Bohrweg 4 2333 CA - Leiden, The Netherlands}
\author{D.J. Thoen}
\affiliation{THz Sensing group, Microelectronics Department, Delft University of Technology, Mekelweg 1, 2628 CD - Delft, The Netherlands}
\author{V. Murugesan}
\affiliation{SRON - Netherlands Institute for Space Research, Niels Bohrweg 4 2333 CA - Leiden, The Netherlands}
\author{N. Llombart}
\affiliation{THz Sensing group, Microelectronics Department, Delft University of Technology, Mekelweg 1, 2628 CD - Delft, The Netherlands}
\author{P. de Visser}
\affiliation{SRON - Netherlands Institute for Space Research, Niels Bohrweg 4 2333 CA - Leiden, The Netherlands}
\affiliation{THz Sensing group, Microelectronics Department, Delft University of Technology, Mekelweg 1, 2628 CD - Delft, The Netherlands}

\date{\today}

\begin{abstract}

Future actively cooled space-borne observatories for the far-infrared, loosely defined as a 1--10 THz band, can potentially reach a sensitivity limited only by background radiation from the Universe. This will result in an increase in observing speed of many orders of magnitude. A spectroscopic instrument on such an observatory requires large arrays of detectors with a sensitivity expressed as a noise equivalent power NEP$\:=\:3\:\times10^{-20}\:\WHz$.
We present the design, fabrication, and characterisation of microwave kinetic inductance detectors (MKIDs) for this frequency range reaching the required sensitivity. The devices are based on thin-film NbTiN resonators which use lens-antenna coupling to a submicron-width aluminium transmission line at the shorted end of the resonator where the radiation is absorbed. We optimised the MKID geometry for a low NEP by using a small aluminium volume of $\approx$ 1$\mum^3$ and fabricating the aluminium section on a very thin (100 nm) SiN membrane. Both methods of optimisation also reduce the effect of excess noise by increasing the responsivity of the  device, which is further increased by reducing the parasitic geometrical inductance of the resonator. 
We measure the sensitivity of eight MKIDs with respect to the power absorbed in the detector using a thermal calibration source filtered in a narrow band around 1.55 THz. We obtain a NEP$_{exp}(P_{abs})\:=\:3.1\pm0.9\times10^{-20}\:\WHz$ at a modulation frequency of 200 Hz averaged over all measured MKIDs. The NEP is limited by quasiparticle trapping. 
The measured sensitivity is sufficient for spectroscopic observations from future, actively cooled space-based observatories. Moreover, the presented device design and assembly can be adapted for frequencies up to $\approx$ 10 THz and can be readily implemented in kilopixel arrays. 
\end{abstract}

\maketitle

\section{Introduction}

Radiation in the far-infrared (FIR) part of the electromagnetic spectrum, loosely defined as the 0.03-1 mm wavelength range, represents about half the energy generated in the Universe since the Big Bang and includes information from processes mostly invisible at other wavelengths \citep{Dole2006}. Unfortunately, observations are notoriously difficult: The Earth's atmosphere is at best partially transparent. Even at the altitudes of stratospheric balloons, line emission from residual water vapour can contaminate astrophysical emission lines. Additionally, self-emission from the telescope creates significant radiation loading, limiting instrument sensitivity. Only an actively cooled space telescope with a temperature of about 4 K in combination with background-limited detectors allows an instrument that is limited only by the Universe background. A spectroscopic instrument on such an observatory requires large-format detector arrays with a pixel count of $\approx10^5$ with a sensitivity expressed in a noise equivalent power (NEP) of $\sim\:$3$\:\times 10^{-20} \WHz$  \citep{Farrah2019,Dunsheath2021}. The combination of sensitivity, high radiation frequency, and pixel count presents a major challenge for future detector systems. Several detector systems exist that are progressing towards this goal. 
\begin{figure*}
\centering
\includegraphics[width=1\textwidth]{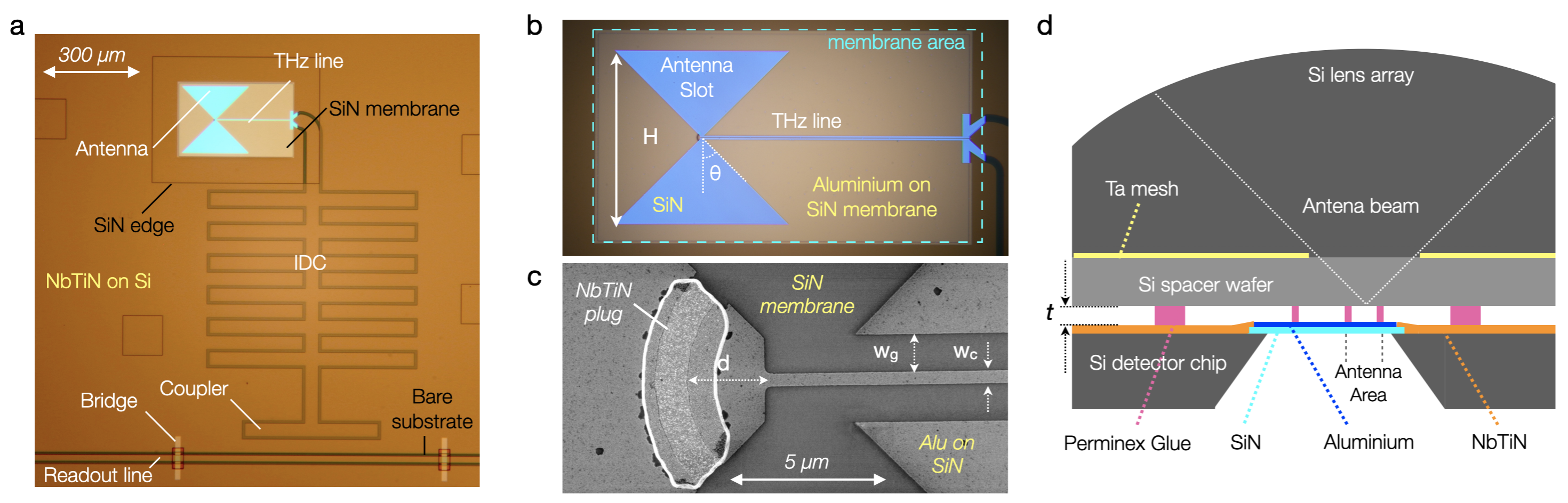}
\caption{\label{Fig:1} Detector geometry. (a) Micrograph of a single detector, consisting of a NbTiN CPW line loaded with an IDC coupled to the readout line via a coupling structure. Its shorted end consists of an aluminium section fabricated on a thin SiN membrane, which is highlighted by the backlighting in the micrograph. (b) Zoom onto the aluminium section on the SiN membrane with the leaky-slot antenna coupled to a co-planar waveguide line (referred to as THz line), where radiation absorption takes place. (c) Scanning electron microscope image of the antenna centre showing the THz line in detail and the NbTiN quasiparticle plug. The THz line dimensions are: ${w_g}$ = 1.2 $\mu$m and ${w_c}$ = 0.4 $\mu$m.  (d) Cross-sectional diagram of the detector assembly (not to scale). The detector chip, as depicted in panels (a-c), is coupled to a Si lens array using a spacer wafer with a Ta absorbing mesh, with an opening aligned to the antenna to enable  radiation coupling. The vacuum gap $t$ is created using spin-on PermiNex\textsuperscript{\textregistered} glue pillars as indicated. }
\end{figure*}

Transition edge sensors (TESs) \citep{Irwin2005} are used in many ground-based and balloon-borne observatories. Recent work \citep{Audley2016,Pourya2016, Williams2020} demonstrates the progress in horn-coupled devices operating in a 60--110 $\mu$m wavelength band reaching a NEP$\:\sim\:$3$\times 10^{-19} \WHz$ using a thermal calibration source \citep{Williams2020}. \citet{Nagler2020} suggest a possible route towards even better sensitivities, reaching an electrical NEP$\:<10^{-20} \WHz$ by means of a proximity-effect TES.

Quantum capacitance detectors (QCDs) \citep{Echternach2013} are a less mature technology, but are the most sensitive: they have shown single-photon counting at 1.5 THz and background limited performance {at} power levels as low as 10$^{-20}$ W, corresponding to a NEP$ \;<\:$10$^{-20}\WHz$ using a 451 pixel array of antenna-coupled detectors operating at 1.55 THz \citep{Echternach2021}.

Microwave kinetic inductance detectors (MKIDs) are microwave resonators whose resonance frequency depends on the amount of radiation absorbed \citep{Day2003}. Antenna-coupled MKIDs have reached sensitivities of NEP = 3$\times 10^{-19} \WHz$ for a 961 pixel array operating around a radiation frequency of 850 GHz, read out by a single readout system \citep{Baselmans2017}. Operation at radiation frequencies above 1 THz with a similar sensitivity and array size was demonstrated by \citet{Bueno2017,Bueno2017kpixel}. Lumped-element KIDs have almost reached similar sensitivities for devices coupled to an on-chip filterbank \citep{Geehan2018}. MKIDs are especially attractive because of their ease of multiplexing, allowing kilopixel arrays with only a single readout line. They are also the only detectors whose sensitivity to ionising radiation has been tested extensively for large arrays with relevant sensitivities \citep{Karatsu2019}, with an extrapolated dead time in L2 of less than 4\% depending on array geometry.  Additionally, \citet{Karatsu2016} found negligible changes in device performance after irradiating the devices with a proton beam simulating the worst-case scenario for radiation absorption of a five-year observation in L2.  

In this paper, we present a MKID design that is scalable for a 0.1--10 THz frequency band and for which we measure  a NEP$_{exp}(P_{abs})\:=\:3.1\pm0.9\times10^{-20}\:\WHz$ at a modulation frequency of 200 Hz averaged over all devices, using a thermal calibration source band-pass filtered around 1.5 THz. Here, $P_{abs}$ is the THz power absorbed in the detector. The design is optimised for low NEP and high responsivity, and will be easily scalable to kilopixel arrays because the readout requirements and fabrication are similar to earlier large system demonstrators \citep{Baselmans2017}.

\section{Device design}
We explain the device design referring to Fig. \ref{Fig:1}: The detector is an antenna-coupled hybrid MKID made from a wide niobium-titanium-nitride (NbTiN) coplanar waveguide (CPW) section loaded with an interdigitated capacitor (IDC) fabricated on a silicon wafer. At the shorted end of the CPW, the MKID consists of an aluminium section on a sillicon-nitride (SiN) membrane of 100 nm in thickness. The latter has a leaky-slot antenna and aluminium CPW line (hereafter referred to as the THz line) to couple to and absorb the THz radiation. The entire structure is a microwave resonator with a resonance frequency ${F_0}\:\approx\:$3 GHz. Radiation absorption in the aluminium THz line modifies the Cooper pair- and quasiparticle density which results in a frequency shift of the resonator read-out using homodyne detection at ${F_0}$ \citep{Day2003}.  Aluminium is chosen as the active material for its superior performance in terms of NEP with respect to other materials.

The first step in optimising the MKID design is to minimise NEP$_{GR}$, which is the fundamental limit in MKID sensitivity, and is given by \citep{Day2003,Visser2014}:
\begin{equation}\label{Eq:GR}
NEP_{GR} = \frac{2\Delta}{\eta_{pb}}\sqrt{\frac{n_{qp}V}{\tau^*_{R}}},
\end{equation}
where $\Delta$ is the energy gap of the superconductor, $n_{qp}$ the density of thermally excited quasiparticles, $\tau^*_{R}$ the experimentally observed quasiparticle recombination time \citep{Flanigan2017}, and $V$ the volume of the  central strip of the THz line. To minimise NEP$_{GR}$, we aim to reduce the volume \citep{Dunsheath2021}, which implies that the THz line must be short while still absorbing all the power coupled in from the antenna. This requires a thin and narrow central line. We use a central conductor width ${w_c\;=\; 0.4 \mum}$ (Fig. \ref{Fig:1}c) fabricated from a sputter-deposited 16 nm thick aluminium film with a sheet resistance of $R_s \;=\;4.1\;\Omega/\square$, corresponding to a resistivity $\rho$ = 6.5$\cdot10^{-8}\:\Omega$m, and critical temperature $T_c\:$= 1.54 K. To match the line to the input impedance of the antenna, $\mathrm{Z_{antenna}\;=\;141+36i \:\Omega}$, obtained from simulations in CST\textsuperscript\textregistered, we use a gap width ${w_g}$ = 1.2 $\mu$m, resulting in a characteristic impedance of the THz line of $\mathrm{Z_{THz\;line}\;=\;194-103i \:\Omega}$. The high imaginary component is caused by the high aluminium sheet resistance. Despite this, we find a good match between the antenna and the THz line, with the reflection co{\"e}fficient ${\mathbf{|}S_\mathrm{11}}\mathbf{|}$ = $-14.5$ dB.  Subsequently we use simulations in SONNET\textsuperscript{\textregistered} based upon the method described by \citet{Endo2020SONNET} to calculate the fraction of the power absorbed in the THz central line with respect to the total power absorbed. We find a high value of 91.5\% caused by the high line impedance. Most of the remaining power, 8.2\%, is absorbed in the ground plane, and the remaining losses are due to SiN absorption and re-radiation. We note that quasiparticles created in the ground plane do not contribute to the detector response because they diffuse away from the THz line. A THz line length of 50 $\mu$m is sufficient to absorb 90\% of the radiation, the minimum  THz line length in our devices is 113 $\mum$. 

We further reduce the aluminium volume by using a single THz line coupled to the antenna, in contrast to the dual THz line design from  \citet{Bueno2017}. To prevent quasiparticle diffusion from the central line to the ground plane, we use a small NbTiN plug creating an Andreev mirror close to the antenna feed point \citep{Andreev1964}, as shown in Fig. \ref{Fig:1}c. The distance $d\:=\:2.5 \:\mum$ in Fig. \ref{Fig:1}c  is motivated by two factors. First, the THz current density, and with that the ohmic losses in the NbTiN plug, strongly depend on the distance $d$ from the start of the THz line, which we simulate in CST\textsuperscript\textregistered. For $d\: >\: 2.5 \:\mum,$ we find that the losses become independent of $d$ and are limited only by losses in the aluminium film. Second, quasiparticle creation is most intense at the start of the narrow section of the THz line where the current density peaks. Initial quasiparticle excitations will relax to the gap edge of aluminium in about $t_{II}$ =1.3 ns \citep{Kozorezov2000}, which corresponds to a diffusion distance $d=\sqrt{\frac{t_{II}}{2\rho N_0 e^2}}  \approx 1.9\;\mu m$, using $N_0$ the single particle density of states in aluminium $N_0 = 1.08\:\cdot 10^{47}\:\mathrm{ m^{-3}J^{-1}}$ and $e$ the electron charge. Hence, $d$ must be larger than this value to prevent quasiparticles created at the start of the narrow section, which have not yet relaxed to the gap energy of the aluminium to escape over the NbTiN Andreev barrier.

A low ${NEP_{GR}}$ alone is not sufficient for the detector to actually reach this sensitivity experimentally. It is also crucial that the measured noise, and thereby the NEP, is dominated by photon fluctuations and not by excess noise sources. We therefore maximise the photon noise power spectral density as measured with an MKID, which is given by \citep{Visser2014}
\begin{equation}\label{Eq:Noise}
S_{\theta}^P = 2P_{abs}hF  \frac{(d\theta/dP_{abs})^2}{1+(2\pi f\tau^*_{R})^2}
\end{equation}
when using the MKID phase $\theta$ as observable, as is the case in this work. Here, we omit the photon bunching term which is negligible for the power and frequency used in this work. The first term represents the photon number fluctuations, with $F$ the photon frequency, $h$ Planck's constant, $f$ the modulation frequency, and $d\theta/dP_{abs}$  the MKID phase responsivity to the absorbed power. It is useful to reformulate this equation in terms of quasiparticle excitations, as an MKID measures photon fluctuations indirectly by measuring the quasiparticles created. Using $dN_{qp}/dP{_{abs}}=\tau^*_R\eta_{pb}/\Delta$, $P{_{abs}}=N_{qp}\Delta/(2\tau^*_R\eta_{pb})$ \citep{Flanigan2017} and the Mattis-Bardeen result for ${d\theta/dn_{qp}}$ \citep{Mattis1958}, we get
\begin{equation}\label{Eq:Noise2}
S_{\theta}^P       = \frac{n_{qp}\tau^*_{R}\eta_{pb}}{1+(2\pi f\tau^*_{R})^2} \frac{hFV}{\Delta}\cdot 
\left(
-\frac{\alpha\beta Q}{|{\sigma}|V}
\frac{d\sigma_2}{dn_{qp}} 
\right)^2
,\end{equation}
where the first two terms express the fluctuations in quasiparticle number due to photon fluctuations and the last term represents the MKID responsivity to quasiparticle number changes, $d\theta/dN_{qp}$ where $N_{qp}=n_{qp}V$ is the total number of quasiparticles. Furthermore, $\beta{\;=1+\frac{2t/\lambda}{\sinh{2t/\lambda}}}\sim\;$2 for thin films, with $\lambda$ the bulk magnetic penetration depth and $t$ the film thickness, $|\sigma|$ the absolute value of the aluminium surface conductance $\sigma=\sigma_1-i\sigma_2$, which depends on the energy gap, quasiparticle density, and microwave frequency \citep{Mattis1958}. $Q$ is the resonator quality factor, $\alpha=L_{k,c}/L_{tot}$, which is the ratio of the kinetic inductance of the THz central line with respect to the total inductance. We maximise $S_{\theta}$ in three ways: (i) We increase both $\tau^*_{R}$ and $\eta_{pb}$ by placing the THz line on a very thin membrane \citep{Fyhrie2016}, in contrast with the 1$\:\mu m$ thick membrane used by \citet{Bueno2017}. This enhances the re-trapping of 2$\Delta$ phonons emitted by quasiparticle recombination. The result is an increase in the phonon escape time $\tau_{esc}$  which increases the experimental quasiparticle recombination time $\tau_R^*=0.5\cdot\tau_R\cdot(1+\tau_{esc}/\tau_B)$. Here $\tau_R$ is the single particle recombination time as defined by \citet{Kaplan1976} and ${\tau_B}$ = 0.26 ns is the pair-breaking time in aluminium \citep{Guruswamy2014}. The membrane thickness is approximately 100 nm, which is limited by its mechanical strength. We find $\tau_{esc}/\tau_B\:$=0.65 using a simple phonon-trapping model \citep{2021deVisser}, which is a factor four increase with respect to a solid substrate. A larger ratio of $\tau_{esc}/\tau_B$ also increases $\eta_{pb}$; we estimate $\eta_{pb}$ = 0.37 based on the work of \citet{Guruswamy2014}, which will be smaller for  the same THz line on a solid substrate. (ii) We minimise the THz central line volume $V$ as discussed before, and (iii) we maximise the kinetic inductance fraction $\alpha$. For the kinetic inductance of the THz line itself, this is automatically achieved by using a narrow and thin THz line. However, $\alpha$ in Eq. \ref{Eq:Noise}  is defined with respect to the entire resonator and a $\lambda/4$ distributed resonator has a significant geometric inductance. To reduce the geometric inductance, we include a large IDC to ground in the MKID as shown in Fig. \ref{Fig:1}a, following \citet{Noorozian2009}. The IDC-section has a much larger capacitance-to-inductance ratio than a CPW line, and therefore the same resonance frequency is obtained with less parasitic inductance in the NbTiN section. The effect is greatest for the shortest THz lines: $\alpha\approx0.6$ for a 113 $\mum$-long THz line, while a device without the IDC would have $\alpha\approx0.42$. This is a relatively small difference because the THz line itself has a very high (kinetic) inductance due to its small dimensions and high sheet resistance.
\begin{figure*}
\centering
\includegraphics[width=1\textwidth]{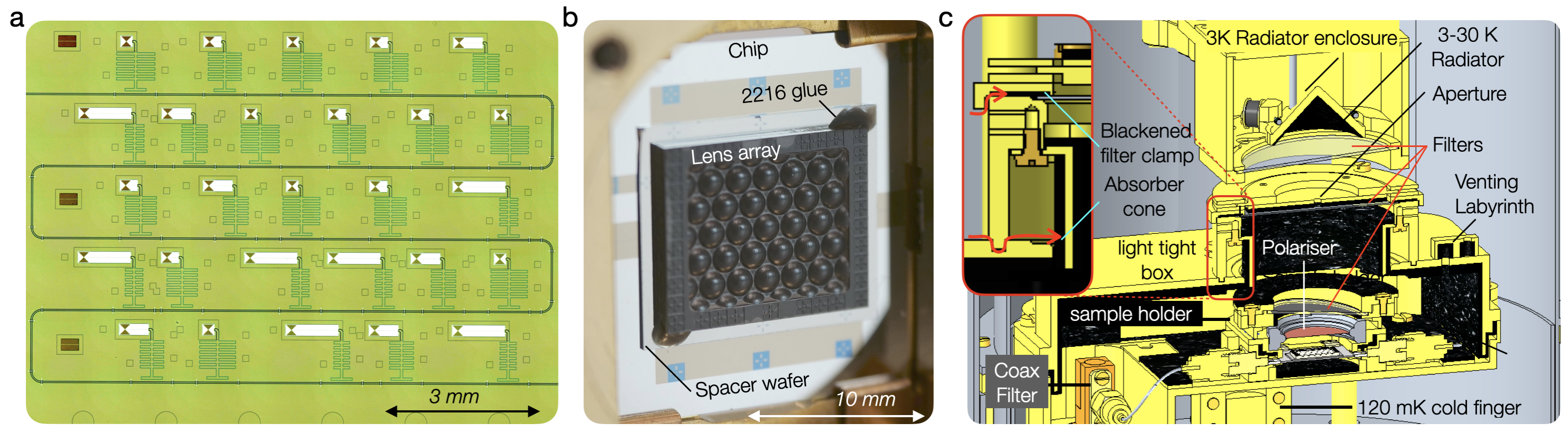}
\caption{\label{Fig:2} Detector chip and experimental setup. (a) A microscope image of the detector chip prior to lens array alignment, showing the 27 different detectors and the four different geometries used, which vary in THz line length. We note that the centre pixel is one of the designs with the shortest THz line length. (b) Photograph of the detector assembly after lens array and spacer wafer mounting. (c) Cross-sectional CAD render of the experimental setup. The upgrades with respect to the setup of \citet{Visser2014} are highlighted in the insert. }
\end{figure*}

The last step in our effort to maximise the photon noise (and GR noise) visibility is to reduce the TLS noise, which we achieve by widening the capacitive section of the resonator, including the IDC \citep{Gao2008b}. We use a central line width $w$ = 40$\mum$ and a gap width of $g\:=\:$8$\mum$. Wider structures become impractically large because of the onset of self-resonances within the IDC itself. Using a narrower gap increases the IDC capacitance, but will increase the TLS noise \citep{Wenner2011}. We therefore limit the ratio $w/(w+2g)$ to about 0.70. We note that an increase in $Q$ does not provide a useful optimisation, because both the photon noise level and the TLS noise level scale with $Q^2$.

For the radiation coupling, we use a leaky wave antenna \citep{Neto2010a} because this antenna allows for relatively large feature sizes, enabling high radiation frequencies \citep{Ozan2016}, and because it must be fabricated on a membrane, compatible with our design optimisation discussed above. The antenna feed is placed in the second focus of a synthesised elliptical Si lens with a diameter of 1.550 mm. The antenna uses a small vacuum gap $t \ll \lambda$ to launch a leaky wave into the Si \citep{Neto2010a}. The antenna design is optimised for 1.5 THz resulting in a slot opening angle of $\theta$ = 45 $^\circ$, a slot length of $H$ = 200 $\mum,$ and a vacuum gap of $t$ = 4 $\mu$m, as shown in Fig. \ref{Fig:1}b,d. The entire design of the KID can be optimised for frequencies up to $\sim$ 10 THz by changing the leaky slot dimensions and vacuum gap height. 

\section{Fabrication and assembly}
The device fabrication is largely identical to that of \citet{Bueno2017} and we use UV lithography and dry etching to define most of the structures. We start with a $\mathrm{\rho\:=\:10\:k\Omega cm}$ Si wafer on which we deposit 150 nm SiN on both sides using low-pressure chemical vapour deposition (LPCVD). The SiN is patterned to define the membrane areas on the  front side of the wafer and etch masks on the back side, which are used in the last fabrication step to remove the Si, creating a free-standing membrane. Subsequently, we sputter deposit 207 nm NbTiN ($\mathrm{T_c\:=\:15\:K\:,\rho\:=\:228\cdot10^{-8}\;\Omega m}$) to define the resonators. The NbTiN etch requires a short over-etch which removes approximately 50 nm from the SiN. To define the THz line and antenna we sputter deposit a 16 nm aluminium film and pattern it in two steps: First we perform UV contact lithography and wet etching to define the readout line bridges and a large patch of aluminium covering the membrane area. We then use electron beam lithography and wet etching to define the final patterns, similarly to \citet{Mirzaei2021}, resulting in a high-resolution definition of the structures as shown in Fig. \ref{Fig:1}c. The small black structures on the aluminium are small voids created during the 15-minute 175 $^\circ$C bake of the PMMA e-beam resist, likely caused by thin-film dewetting \citep{Thompson2012}. In the last step, we open the membranes by potassium-hydroxide (KOH) etching  (8 hours in 75$^\circ$C KOH:H$_{\mathrm{2}}$O 1:4), followed directly with a RCA2 clean (10 minutes in 70$^\circ$C HCl:H$_\mathrm{2}$O$_\mathrm{2}$:H$_\mathrm{2}$O 1:1:5) to remove potassium ions on the back side of the membrane. 
\begin{figure*}
\centering
\includegraphics[width=1\textwidth]{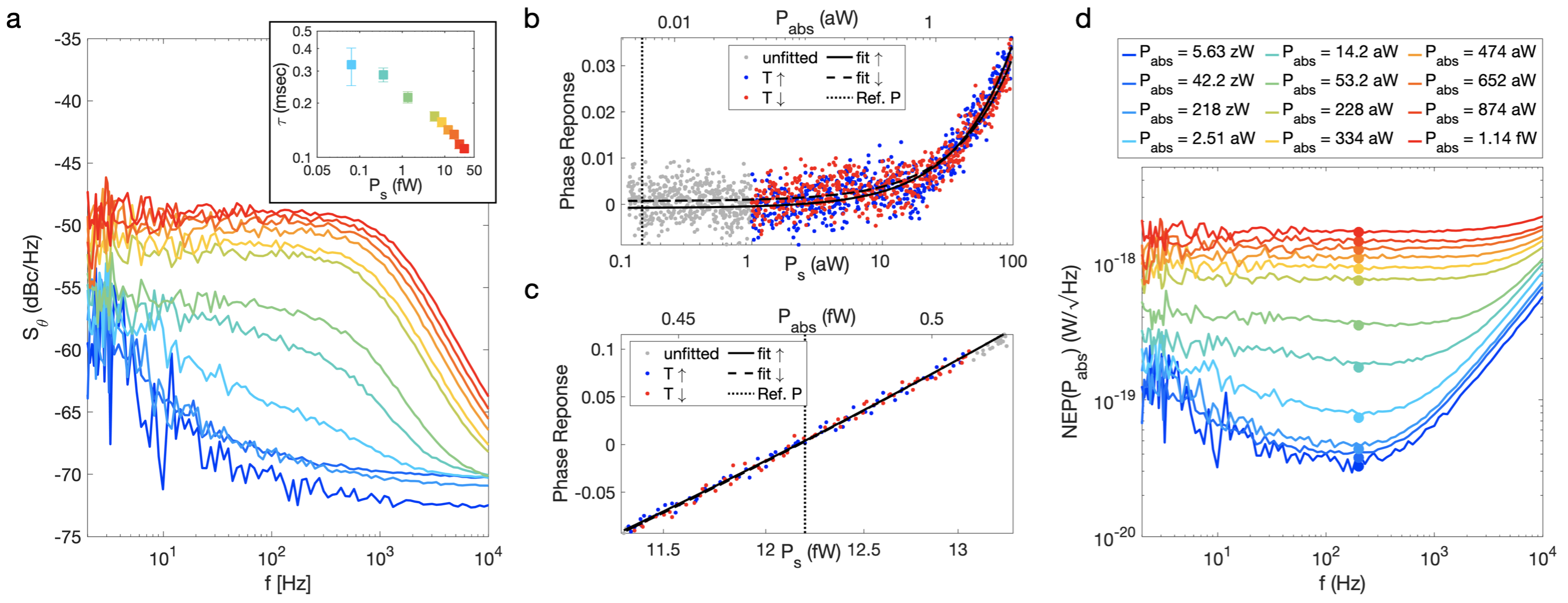}
\caption{\label{Fig:3} Noise, response and NEP for a single detector. (a) Measured phase noise power spectral density for several source powers, increasing from bottom to top. There is a clear transition from detector-limited noise (lower curves) to photon-noise-limited noise, characterised by a white spectrum rolled-off with $\tau^*_R$ (highest curves).  The insert shows the $\tau^*_R$ for the same source powers. (b) KID phase response around the lowest source power used to obtain the NEP, $P_s$ = 0.145 aW ($P_{abs}$ = 5.63 zW), which corresponds to a radiator temperature of 4 K. The data represent a measurement obtained by sweeping the radiator temperature from 3.95 K to 6.21 K and back to 3.95 K. The data is plotted on a semi-logarithmic scale to show the absence of any response at low powers. The fits and fitted range are indicated as well. (c) MKID response and fits for increasing and decreasing radiator temperature around a higher source power, $P_s$ = 12.2 fW. We indicate the absorbed power as well as the absorbed power in panels (b) and (c) for convenience. (d) Measured NEP for the same set of powers, now expressed in power absorbed in the detector. The legend in absorbed power is also valid for panel (a). } 
\end{figure*}

The assembled detector array consists of the device chip, a 250 $\mum$ Si spacer wafer, and a laser-machined Si lens array commercially obtained from Veldlaser. A  cross-sectional sketch of the detector assembly is given in Fig. \ref{Fig:1}d, and a photograph is shown in Fig. \ref{Fig:2}b. The spacer wafer top has a 40 nm $\beta$-Ta absorbing mesh $(R_s\:=\:\mathrm{51\:\Omega/\square}$) with a design similar to that used by \citet{Baselmans2017} that acts as a low-pass filter at the MKID readout frequencies and as THz radiation absorber to eliminate stray radiation. Above each antenna, a 600 $\mu$m diameter hole allows radiation coupling to the antenna. On the other side, the spacer wafer has 4 $\mum$ thick lithographically defined pillars made from PermiNex\textsuperscript{\textregistered}, and a spin-on photo-sensitive glue used to define the vacuum gap $t$ as indicated in Fig. \ref{Fig:1}d and to provide a permanent bond between the chip and the spacer wafer. To assemble the final detector, we align the spacer wafer with the lens array and glue them together with two spots of 2216 epoxy (see Fig. \ref{Fig:2}b), which is cured for 30 minutes at 90$^\circ$C. At this temperature the PermiNex\textsuperscript{\textregistered} is unaffected. In the next step, we align the lens-spacer assembly to the chip and press the assembly together using a spring-loaded tool. The PermiNex\textsuperscript{\textregistered} is cured for 15 minutes at 180 $^\circ$C to create a permanent bond between spacer and chip, which is found to be resilient against more than 20 thermal cycles down to 120 mK. A photograph of the entire detector chip without the lens array is given in Fig. \ref{Fig:2}a. The chip has 27 detectors, with four different device designs that differ in THz line length, which can be 913, 663, 238, or 113 $\mum$. This corresponds to $V$ = 6.0, 4.4, 1.6, and 0.85 $\mum^3$ respectively, including the small volume near the NbTiN plug. All devices are designed with a coupling Q factor  $Q_{c}$=120$\cdot\mathrm{10^3}$. 

\section{Experiment}
In the experiment, we measure the sensitivity, NEP$_{exp}(P_s,f)$, at several powers from a thermal radiator source $P_s$ using the setup depicted in Fig. \ref{Fig:2}. We denote with $f$ the post-detection modulation frequency of the spectral shape of the NEP. We mount the detector assembly in a holder which itself is mounted inside a light-tight box mechanically anchored to the cold stage of an adiabatic demagnetisation refrigerator (ADR). All interfaces and venting holes of the holder and light-tight box are equipped with labyrinths coated with a radiation-absorbing layer from carbon-loaded epoxy and SiC grains. A thermal radiation source weakly coupled to the 3 K stage of the cooler enclosed in a 3 K box is placed directly above, and can be heated to 30 K. In the first experiment, we used a 24 mm diameter radiator with a radiation absorbing layer from carbon-loaded epoxy and 0.5 mm SiC grains, as shown in Fig. \ref{Fig:2}c. Radiation coupling between the source and the detector is achieved through a 1.85 mm aperture at 42 mm from the lens surface. The use of the aperture reduces the throughput to the detectors, and therefore higher temperatures are needed to reach the same amount of source power at the detectors. This reduces the sensitivity of the radiation power to errors in the measured temperature. Furthermore, it allows a check as to whether the coupled power matches the beam pattern of the detector. We use eight filters in total ---mounted on the radiator box, the light-tight box, and the sample holder--- to define a pass-band of around 1.55 THz. These are the exact same filters as discussed by  \citet{Visser2014}. The second experiment was identical, except that we used a 40 mm diameter radiator with 1 mm SiC grains and no aperture, to cross-check our results: In this case, the antenna pattern does not play a role because the opening angle to the radiator is much larger than the beam opening angle. Also, the total power coupled to the detector chip largely exceeds the power coupled to the single-mode antenna beams. Stray coupling to this excess radiation will result in overestimation of the coupling efficiency.  In all cases, a polariser is used on the holder to allow only radiation in co-polarisation with respect to the antenna. We note that, because of the increased
sensitivity of the detectors, we needed to upgrade the light-tight box with respect to the setup in  \citet{Visser2014} ---as indicated in the insert of Fig. \ref{Fig:2}c--- with better stray-light control around the filter mount. We use a CRYOPHY\textsuperscript{\textregistered} outer shield and niobium inner shield at 3K surrounding the light-tight box and radiator to limit time-dependent magnetic fields. The theoretical coupling between the absorbing aluminium volume in the MKID aligned with the aperture is calculated using a CST simulation of the antenna beam and the tool provided by \citet{Huasheng2019} to be $\eta_{opt, calc.}\:=\:$4.3\%. The reason for the low coupling is the (intended) large spill-over loss between the small pin-hole aperture and the lens--antenna beam patterns. This includes the losses in the ground plane of the THz line and the antenna to THz line mismatch, which are discussed above. For the 40 mm diameter radiator without aperture, we find $\eta_{opt, calc.}\:=\:$76\%.

We began experiments with a frequency sweep with the chip at 120 mK and the radiator at 3 K, which represents fully dark conditions. We find 23 out of 27 MKIDs with an average $Q_i=0.23\cdot10^6\pm0.13\cdot10^6$ and an average $Q_c=6\cdot10^4\pm2\cdot10^4$ at a readout power of $-118$ dBm. We selected a subset of the detectors that couple well to the aperture. For each detector we obtained NEP$_{exp}(P_s,f)$ using
\begin{equation}\label{Eq:NEPexp}
NEP_{exp}(P,f) = \sqrt{S_{\theta}(f)}\left(\frac{d\theta}{dP}\right)^{-1}\sqrt{1+(2\pi f\tau_R^*)^2},
\end{equation}
where $P$ is the radiation power, given now by $P_s$, which is the power from the source that can couple to a single mode. ${P_s}$ is calculated from the radiator temperature and the measured filter transmission using the formalism explained in detail in Appendix B of \citet{Ferrari2018}.  Figure \ref{Fig:3}a shows the phase noise power spectral density $S_{\theta}(f)$ for one representative device, which has a resonance frequency of 2.74 GHz and an aluminium length of 913 $\mum$. The spectrum is obtained from 64 seconds of time domain data sampled at 50 kilosample/sec with the readout tuned to the MKID resonance frequency while maintaining a constant black body temperature. At low powers, shown by the lower (blue) lines, we observe a noise spectrum with a 1/f slope. At increasing power, the noise level at frequencies f $<$ 1 kHz increases sharply until the noise becomes white for $P_s\:>$ 1 fW.  The insert in the figure shows $\tau_R^*$ for the same powers obtained from a fit to the cross power spectral density of the resonator amplitude and phase noise \citep{Visser2012b}, which is given in Appendix \ref{Appendix1}. We observe a saturation of the recombination time at around 0.3 ms and a reduction for increasing powers as expected. Subsequently, we obtain the responsivity  $d\theta/dP_s$ by a linear fit to the MKID phase response upon a small change of the black-body temperature. Care was taken to sweep the radiator temperature slowly to prevent hysteresis caused by a temperature difference between the thermometer and the radiating surface. In Fig. \ref{Fig:3}b, we show the response measurement and fit around the lowest $P_s$ of 0.145 aW (corresponding to a 4 K radiator temperature). We fit the response curve over a range of 1 aW $<\:P_s\:<$ 0.1 fW (which corresponds to a range in absorbed power 0.039 aW $<\:P_{abs}\:<$ 39 aW, as we show later): the high power limit is used so as to reach a statistically significant response which is still linear; the low end was chosen so as to reduce the sensitivity to drifts. It is clear from the
figure that there is negligible hysteresis and that the fits are excellent over the whole power range spanning a factor 1000, with no response below $P_s$ = 1 aW. This proves that the setup is light tight; radiation leaks would allow unfiltered radiation to couple to the detector. The total integrated power per mode for a 4 K radiator is $10^6$ times larger than the power in the narrow band around 1.5 THz, and it is strongly temperature dependent. Even a weak coupling to broad-band stray radiation would therefore result in a significant response of the detector at 4 K ($P_s$ = 0.145 aW) in Fig. \ref{Fig:3}b. For higher powers, we fit a symmetric power range around the power at which we measure the noise, as shown in Fig. \ref{Fig:3}c. 

A possible source of systematic error is the radiator temperature readout: The detector response starts at $P_s$ = 1 aW, corresponding to a radiator temperature of 6.3 K, where a 50 mK temperature change would result in a 10\% change in $P_s$. For the full aperture setup, the response starts at a temperature of 4.5 K, which is due to the higher coupling. At this temperature, a 50 mK temperature change will result in a 20\% change in $P_s$. We therefore performed a cross-calibration between the Cernox\textsuperscript{\textregistered} radiator thermometer and another similar thermometer. The temperature difference between the two is less than 15 mK in the temperature range of the experiment, contributing less than 3\% uncertainty in radiator power. 

\begin{figure*}
\centering
\includegraphics[width=1\textwidth]{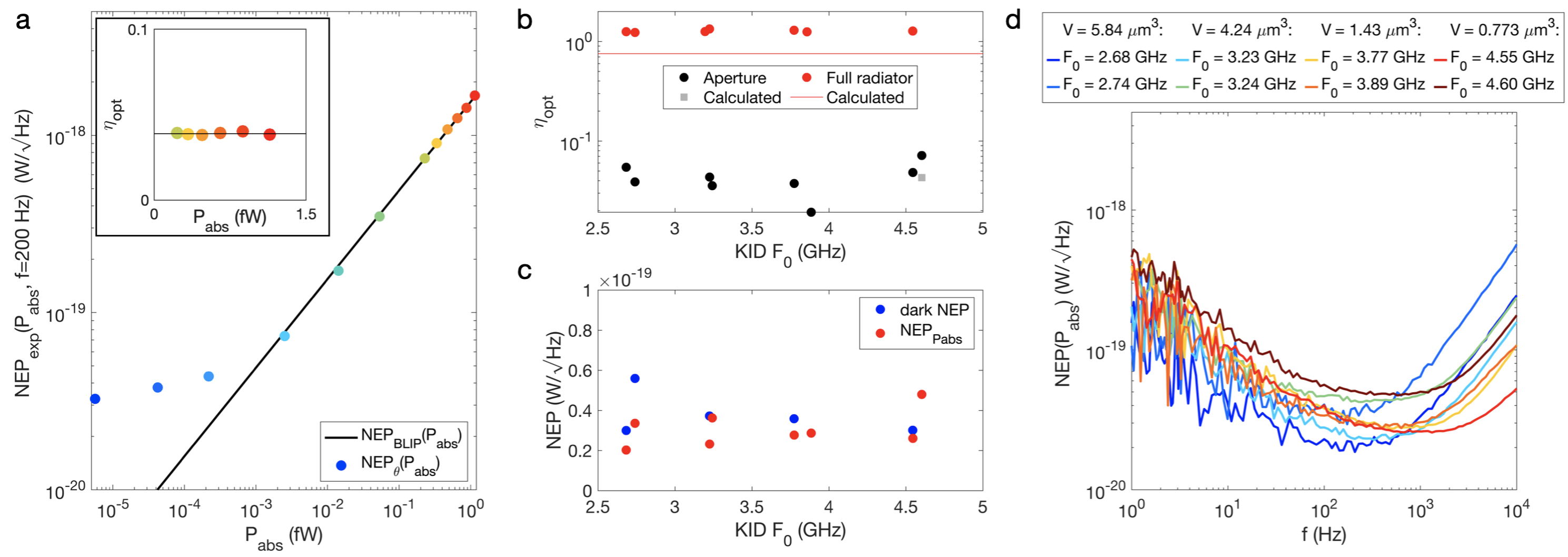}
\caption{\label{Fig:4} NEP and coupling efficiency. (a) The NEP at 200 Hz, as indicated by the dots in panel (d), as function of absorbed power using the pinhole setup for the same devices as in Fig.\ref{Fig:3}. The black line shows the theoretical curve. The absorbed power is evaluated from the source power using the measured efficiency as indicated in the insert, and the mean value of the points, $\eta_{opt}$ = 0.0389 $\pm$ 0.004, is given by the line. The statistical error bars are smaller than the dots. (b) The measured NEP$_{exp}(P_{abs},\:f\;\mathrm{\:=\:200 Hz})$ for the pinhole experiment, as displayed in Fig. \ref{Fig:2}c, together with the dark NEP obtained at the same modulation frequency using a change in bath temperature to obtain the response. (c) The averaged coupling efficiencies for the two experiments. The red dots represent the results for the large radiator, where all detectors couple almost identically to the source as expected; the expected coupling is given by the line. The black dots give the coupling for the pinhole setup, where only the best aligned detector is expected to reach the theoretical coupling value given by the black line. We note that, in both cases, the measured coupling exceeds the theoretical value by the same factor 1.6. (d) The measured NEP$_{exp}(P_{abs},\:f)$ for the pinhole experiment at the lowest radiator temperature for all measured MKIDs. We observe no dependence of the NEP on aluminium volume. We note that panel (a) and Fig.\ref{Fig:3} are given for the detector with F$\mathrm{_0}$ = 2.74 GHz.}
\end{figure*}

We can now obtain NEP$_{exp}$ using Eq. \ref{Eq:NEPexp}, and use it to determine the detector coupling efficiency $\eta_{opt}$ using the photon noise from the radiator as an absolute calibration source \citep{Ferrari2018}. This procedure is only valid if the MKID is photon-noise limited, which is the case for source powers exceeding 5 fW; here the spectrum is white with a single Lorentzian roll-off \citep{Janssen2013}. The optical efficiency is given by \citep{Ferrari2018}
\begin{equation}\label{Eq:eta}
\eta_{opt}= \frac{\int{2P_{s,\nu} h \nu d\nu} + \int{4\Delta P_{s,\nu}/\eta_{pb}}d\nu }    {NEP_{exp}^2 - \int{2P_{s,\nu}h\nu F_{\nu}O_\nu d\nu}}
,\end{equation}
where $F_\nu$ is the filter transmission, $O_\nu$ is the photon occupation number, and $P_{s,\nu}$ is $P_s$ per unit frequency $\nu$. The result is given in the insert of Fig. \ref{Fig:4}a. We get $\eta_{opt}\:\mathrm{= 0.039}\:\pm\:$0.004 with very little scatter between the individual datapoints. We can now obtain the experimental NEP as a function of the power absorbed, NEP$_{exp}(P_{abs},f),$ using Eq. \ref{Eq:NEPexp} with $P=P_{abs}=\eta_{opt}P_s$. The result is shown in Fig. \ref{Fig:3}d. At the lowest powers, the NEP at $f$ = 200 Hz reaches a value of NEP$_{exp}(P_{abs}f\:\mathrm{=\:200\:Hz})\;=\;3.3\pm0.3\times10^{-20}\; \WHz$. At lower frequencies, the NEP is higher because of the 1/f noise. For higher powers, the NEP spectral shape becomes flatter, and for $P_{abs}\:>\:$53 aW the NEP is frequency independent down to 1 Hz. In Fig. \ref{Fig:4}a, we give NEP$_{exp}(P_{abs},f\:\mathrm{=\:200\:Hz})$ together with the theoretical value of the photon noise background from the radiation source and obtain a result very similar to those of \citet{Visser2014},\citet{Baselmans2017} and \citet{Ferrari2018}, but with ten times greater sensitivity. For a comparison of the NEP as a function of the absorbed power as shown in Fig. \ref{Fig:4}a for all measured MKIDs, we refer to Appendix \ref{Appendix3}. For all MKIDs, the microwave readout power absorbed by the quasiparticle system, $P_{abs,qp}$, is significant, especially at low absorbed powers: For example, at the lowest value of $P_{abs}$ = 5.63 zW, we estimate that $P_{abs,qp}\;\simeq$ 20 aW. At values of $P_{abs}\;\geq$ 1 fW, $P_{abs,qp}$ and $P_{abs}$ become similar.

In Fig. \ref{Fig:4}b, we give the coupling efficiency for all MKIDs for the aperture-coupled radiator by the black dots, and the theoretical coupling for a well-aligned detector by the grey square. We observe that the MKID with the highest resonance frequency has the maximum coupling, as expected as this is the detector in the chip centre as can be seen in Fig. \ref{Fig:2}a. The measured $\eta_{opt}\:=\:$7\%, exceeding the expected value of 4.3\% by a factor 1.63. We observe six MKIDs with similar coupling but some scatter, which can be explained by a small misalignment of the aperture, causing unequal coupling to the six detectors at equal distance from the centre device. The red dots represent the results with the large radiator; we now see that all detectors couple identically with a mean value of  $\eta_{opt}\:=\:1.27\;\pm\:0.04$, exceeding the expected value by a factor 1.68. The fact that both experiments yield an identical coupling with respect to the theoretical value implies that both setups are equivalent and systematic errors such as an error in the beam pattern calculation, in-band stray radiation coupling with the large radiator, or a difference in temperature between the radiator surface and thermometer can be excluded. The higher measured transmission is therefore most likely caused by a higher transmission of our eight quasi-optical filters compared to the data provided by their supplier QMCI\textsuperscript{\textregistered}, which gives a peak transmission of 13.5\%. A potential reason is that QMCI\textsuperscript{\textregistered} performs the filter measurements at room temperature.

In Fig. \ref{Fig:4}c we give NEP$_{exp}(P_{abs},\:f\:\mathrm{\:=\:200 Hz})$ for eight KIDs in the pinhole experiment and find a mean value of NEP$_{exp}(P_{abs},\:f\mathrm{\:=\:200 Hz})\:=\:3.1\pm0.9\times10^{-20}\:\WHz$. We also give the dark NEP \citep{Baselmans2008,Janssen2014b}. To obtain the dark NEP, we measure the response of the MKIDs to a change in chip temperature while keeping the radiator in dark conditions, that is, $\mathrm{T_{BB}}$ = 3 K. Under these conditions, the number of quasiparticles in the aluminium $\mathrm{N_{qp}}$ can be calculated from the chip temperature, the volume of the aluminium section of the resonator, and the energy gap \citep{Janssen2014b}. Rewriting $\mathrm{N_{qp}}$ in terms of FIR power allows the dark NEP to be calculated using Eq. \ref{Eq:NEPexp} by replacing the responsivity term $d\theta/\delta P$  by the dark responsivity given by  \citet{Baselmans2008}:
\begin{equation}
\label{Eqn:NEPdark}
\frac{d\theta}{d P_{dark}} = \frac{\eta_{pb} \tau^*_R}{\Delta}\frac{d\theta}{dN_{qp}(T)}
,\end{equation}
where we obtain $\tau^*_R$ using a fit to the cross power spectral density of the phase- and amplitude noise as discussed above. For the operating temperature of 120 mK, a good fit cannot always be obtained. We find a mean value of NEP$_{{dark}},\:f\mathrm{\:=\:200 Hz})\:=\:3.8\pm1.3\times10^{-20}\:\WHz$, in good agreement with NEP$_{exp}(P_{abs}\:f\mathrm{\:=\:200 Hz})$. A similar agreement between dark and optical NEP was found by \citet{Janssen2014b}. Finally, panel (d) shows the NEP$_{exp}(P_{abs},f)$ at the lowest radiator power for the same set of MKIDs. All have a similar sensitivity. 

\section{Discussion}
The observed reduction in photon noise level for $P_{abs}\:<\:$228 aW (see Fig. \ref{Fig:3}a) is inconsistent with a NEP limited by a thermal reservoir of residual quasiparticles, which was the conclusion in previous work \citep{Visser2014}. However, it is consistent with quasiparticle trapping as discussed by \citet{2021deRooij}. The optimisation of the MKID responsivity, as discussed in the context of Eq. \ref{Eq:Noise2}, is now especially relevant. A high photon noise level allows for the device to remain background limited even when the photon-noise level drops. The use of a thin membrane, increasing both $\tau^*_R$ and $\eta_{pb}$, is therefore an important optimisation. The downside is a strong reduction in instantaneous dynamic range when compared to the results of \citet{Baselmans2017}. We find an instantaneous dynamic range of a few hundred aW, with readout re-tuning the devices operate up to 30 fW absorbed power. 

Interestingly, we observe in Fig. \ref{Fig:4} that the NEP$_{exp}(P_{abs},\:f\mathrm{=\:200\:Hz})$ is independent of the aluminium volume. More precisely, we observe that both the noise $S_{\theta}$ and responsivity $d\theta/dP$ are independent of volume. For the noise, this is not surprising, because we are limited in this regime by amplifier noise and device noise, but the responsivity is expected to scale with the inverse of the volume, as can be seen from Eqs. \ref{Eq:Noise} and \ref{Eq:Noise2}. We find from the dark NEP measurement that $d\theta/dN_{qp}$ {does} scale inversely with volume as expected. We also find that the low temperature quasiparticle lifetime $\tau^*_R$ scales inversely with volume. A smaller effect is that $\alpha$  is slightly
smaller for the smallest volume devices due to parasitic inductance in the IDC, and the net result is the observed volume independence of the responsivity to power. Interestingly, the observed volume (or line length length) dependence in $\tau^*_R$ is not seen in the Appendix of \citet{2021deRooij}.

A better understanding of the trapping mechanism and the transition between trapping-dominated hybrid MKIDs and devices limited by quasiparticle creation due to the readout power will be important for further optimisation, which is especially relevant for the 1/f noise. If trapping is caused by surface states, as argued by \citet{2021deRooij}, device optimisation is possible using thicker aluminium films as this will reduce trapping and increase $\tau^*_R$. Interestingly, thicker films will also increase $\tau^*_R$ as well as $\eta_{pb}$ due to enhanced $2\Delta$ phonon recycling caused by the longer dwell time of the phonons in the film. This might partially compensate the responsivity reduction caused by an increase in the volume and reduction in $\alpha$. 

It is interesting to consider the intrinsic energy resolution for the possibility of single photon counting in the FIR. This resolution is given by \citet{Moseley1984}: $dE\simeq NEP_{exp}(P_{abs},f\:=\:\mathrm{200\:Hz})\cdot\sqrt{\tau^*_R}$. We find an energy resolution averaged over all MKIDs of $dE/h $ = 0.5 $\pm$ 0.2 THz, with a best value of $dE/h $ = 0.24 THz for one of the devices with the smallest volume. Reliable single photon detection should therefore be possible for frequencies above a few THz. For more details, we refer to Appendix \ref{Appendix2}.

\section{Conclusions}
In conclusion, we present a new MKID design, where radiation coupling and absorption take place in a small aluminium volume fabricated on a 100 nm SiN membrane embedded in a NbTiN resonator. The device is optimised for high quasiparticle responsivity and low NEP. We obtain a mean NEP$_{exp}(P_{abs},f\:=\:\mathrm{200\:Hz})\:=\:3.1\pm0.9\times10^{-20}\:\WHz$ with a time constant varying from 0.33 msec to 0.03 msec. The NEP is limited by 1/f noise and quasiparticle trapping, limiting the performance in dark conditions at low modulation frequencies. At absorbed powers of 53 aW and higher the device reaches background-limited operation with a flat noise level down to 1 Hz. The antenna design, device geometry, and assembly using PermiNex\textsuperscript{\textregistered}  spin-on glue allows scaling to frequencies up to $\approx$ 10 THz. Importantly, creating kilopixel arrays from the presented device design is straightforward; the fabrication and readout are largely identical to those in the work presented by  \citet{Bueno2017kpixel}. All presented data, measurement data, and analysis scripts will be made available in \citet{2022Zenodo}

\acknowledgements{P.J.dV. is financially supported by the Netherlands Organisation for Scientific Research NWO (Veni Grant 639.041.750 and Projectruimte 680-91-127).}

\section{References}
\bibliography{Jochem_biblio20220118}

\begin{thebibliography}{43}%
\makeatletter
\providecommand \@ifxundefined [1]{%
 \@ifx{#1\undefined}
}%
\providecommand \@ifnum [1]{%
 \ifnum #1\expandafter \@firstoftwo
 \else \expandafter \@secondoftwo
 \fi
}%
\providecommand \@ifx [1]{%
 \ifx #1\expandafter \@firstoftwo
 \else \expandafter \@secondoftwo
 \fi
}%
\providecommand \natexlab [1]{#1}%
\providecommand \enquote  [1]{``#1''}%
\providecommand \bibnamefont  [1]{#1}%
\providecommand \bibfnamefont [1]{#1}%
\providecommand \citenamefont [1]{#1}%
\providecommand \href@noop [0]{\@secondoftwo}%
\providecommand \href [0]{\begingroup \@sanitize@url \@href}%
\providecommand \@href[1]{\@@startlink{#1}\@@href}%
\providecommand \@@href[1]{\endgroup#1\@@endlink}%
\providecommand \@sanitize@url [0]{\catcode `\\12\catcode `\$12\catcode
  `\&12\catcode `\#12\catcode `\^12\catcode `\_12\catcode `\%12\relax}%
\providecommand \@@startlink[1]{}%
\providecommand \@@endlink[0]{}%
\providecommand \url  [0]{\begingroup\@sanitize@url \@url }%
\providecommand \@url [1]{\endgroup\@href {#1}{\urlprefix }}%
\providecommand \urlprefix  [0]{URL }%
\providecommand \Eprint [0]{\href }%
\providecommand \doibase [0]{http://dx.doi.org/}%
\providecommand \selectlanguage [0]{\@gobble}%
\providecommand \bibinfo  [0]{\@secondoftwo}%
\providecommand \bibfield  [0]{\@secondoftwo}%
\providecommand \translation [1]{[#1]}%
\providecommand \BibitemOpen [0]{}%
\providecommand \bibitemStop [0]{}%
\providecommand \bibitemNoStop [0]{.\EOS\space}%
\providecommand \EOS [0]{\spacefactor3000\relax}%
\providecommand \BibitemShut  [1]{\csname bibitem#1\endcsname}%
\let\auto@bib@innerbib\@empty
\bibitem [{\citenamefont {{Dole}}\ \emph {et~al.}(2006)\citenamefont {{Dole}},
  \citenamefont {{Lagache}}, \citenamefont {{Puget}}, \citenamefont {{Caputi}},
  \citenamefont {{Fern{\'a}ndez-Conde}}, \citenamefont {{Le Floc'h}},
  \citenamefont {{Papovich}}, \citenamefont {{P{\'e}rez-Gonz{\'a}lez}},
  \citenamefont {{Rieke}},\ and\ \citenamefont {{Blaylock}}}]{Dole2006}%
  \BibitemOpen
  \bibfield  {author} {\bibinfo {author} {\bibfnamefont {H.}~\bibnamefont
  {{Dole}}}, \bibinfo {author} {\bibfnamefont {G.}~\bibnamefont {{Lagache}}},
  \bibinfo {author} {\bibfnamefont {J.-L.}\ \bibnamefont {{Puget}}}, \bibinfo
  {author} {\bibfnamefont {K.~I.}\ \bibnamefont {{Caputi}}}, \bibinfo {author}
  {\bibfnamefont {N.}~\bibnamefont {{Fern{\'a}ndez-Conde}}}, \bibinfo {author}
  {\bibfnamefont {E.}~\bibnamefont {{Le Floc'h}}}, \bibinfo {author}
  {\bibfnamefont {C.}~\bibnamefont {{Papovich}}}, \bibinfo {author}
  {\bibfnamefont {P.~G.}\ \bibnamefont {{P{\'e}rez-Gonz{\'a}lez}}}, \bibinfo
  {author} {\bibfnamefont {G.~H.}\ \bibnamefont {{Rieke}}}, \ and\ \bibinfo
  {author} {\bibfnamefont {M.}~\bibnamefont {{Blaylock}}},\ }\bibfield  {title}
  {\enquote {\bibinfo {title} {{The cosmic infrared background resolved by
  Spitzer. Contributions of mid-infrared galaxies to the far-infrared
  background}},}\ }\href {\doibase 10.1051/0004-6361:20054446} {\bibfield
  {journal} {\bibinfo  {journal} {Astronomy and Astrophysics}\ }\textbf
  {\bibinfo {volume} {451}},\ \bibinfo {pages} {417--429} (\bibinfo {year}
  {2006})},\ \Eprint {http://arxiv.org/abs/astro-ph/0603208} {astro-ph/0603208}
  \BibitemShut {NoStop}%
\bibitem [{\citenamefont {{Farrah}}\ \emph {et~al.}(2019)\citenamefont
  {{Farrah}}, \citenamefont {{Smith}}, \citenamefont {{Ardila}}, \citenamefont
  {{Bradford}}, \citenamefont {{Dipirro}}, \citenamefont {{Ferkinhoff}},
  \citenamefont {{Glenn}}, \citenamefont {{Goldsmith}}, \citenamefont
  {{Leisawitz}}, \citenamefont {{Nikola}}, \citenamefont {{Rangwala}},
  \citenamefont {{Rinehart}}, \citenamefont {{Staguhn}}, \citenamefont
  {{Zemcov}}, \citenamefont {{Zmuidzinas}}, \citenamefont {{Bartlett}},
  \citenamefont {{Carey}}, \citenamefont {{Fischer}}, \citenamefont
  {{Kamenetzky}}, \citenamefont {{Kartaltepe}}, \citenamefont {{Lacy}},
  \citenamefont {{Lis}}, \citenamefont {{Locke}}, \citenamefont
  {{Lopez-Rodriguez}}, \citenamefont {{MacGregor}}, \citenamefont {{Mills}},
  \citenamefont {{Moseley}}, \citenamefont {{Murphy}}, \citenamefont
  {{Rhodes}}, \citenamefont {{Richter}}, \citenamefont {{Rigopoulou}},
  \citenamefont {{Sanders}}, \citenamefont {{Sankrit}}, \citenamefont
  {{Savini}}, \citenamefont {{Smith}},\ and\ \citenamefont
  {{Stierwalt}}}]{Farrah2019}%
  \BibitemOpen
  \bibfield  {author} {\bibinfo {author} {\bibfnamefont {D.}~\bibnamefont
  {{Farrah}}}, \bibinfo {author} {\bibfnamefont {K.~E.}\ \bibnamefont
  {{Smith}}}, \bibinfo {author} {\bibfnamefont {D.}~\bibnamefont {{Ardila}}},
  \bibinfo {author} {\bibfnamefont {C.~M.}\ \bibnamefont {{Bradford}}},
  \bibinfo {author} {\bibfnamefont {M.}~\bibnamefont {{Dipirro}}}, \bibinfo
  {author} {\bibfnamefont {C.}~\bibnamefont {{Ferkinhoff}}}, \bibinfo {author}
  {\bibfnamefont {J.}~\bibnamefont {{Glenn}}}, \bibinfo {author} {\bibfnamefont
  {P.}~\bibnamefont {{Goldsmith}}}, \bibinfo {author} {\bibfnamefont
  {D.}~\bibnamefont {{Leisawitz}}}, \bibinfo {author} {\bibfnamefont
  {T.}~\bibnamefont {{Nikola}}}, \bibinfo {author} {\bibfnamefont
  {N.}~\bibnamefont {{Rangwala}}}, \bibinfo {author} {\bibfnamefont {S.~A.}\
  \bibnamefont {{Rinehart}}}, \bibinfo {author} {\bibfnamefont
  {J.}~\bibnamefont {{Staguhn}}}, \bibinfo {author} {\bibfnamefont
  {M.}~\bibnamefont {{Zemcov}}}, \bibinfo {author} {\bibfnamefont
  {J.}~\bibnamefont {{Zmuidzinas}}}, \bibinfo {author} {\bibfnamefont
  {J.}~\bibnamefont {{Bartlett}}}, \bibinfo {author} {\bibfnamefont
  {S.}~\bibnamefont {{Carey}}}, \bibinfo {author} {\bibfnamefont {W.~J.}\
  \bibnamefont {{Fischer}}}, \bibinfo {author} {\bibfnamefont {J.}~\bibnamefont
  {{Kamenetzky}}}, \bibinfo {author} {\bibfnamefont {J.}~\bibnamefont
  {{Kartaltepe}}}, \bibinfo {author} {\bibfnamefont {M.}~\bibnamefont
  {{Lacy}}}, \bibinfo {author} {\bibfnamefont {D.~C.}\ \bibnamefont {{Lis}}},
  \bibinfo {author} {\bibfnamefont {L.}~\bibnamefont {{Locke}}}, \bibinfo
  {author} {\bibfnamefont {E.}~\bibnamefont {{Lopez-Rodriguez}}}, \bibinfo
  {author} {\bibfnamefont {M.}~\bibnamefont {{MacGregor}}}, \bibinfo {author}
  {\bibfnamefont {E.}~\bibnamefont {{Mills}}}, \bibinfo {author} {\bibfnamefont
  {S.~H.}\ \bibnamefont {{Moseley}}}, \bibinfo {author} {\bibfnamefont {E.~J.}\
  \bibnamefont {{Murphy}}}, \bibinfo {author} {\bibfnamefont {A.}~\bibnamefont
  {{Rhodes}}}, \bibinfo {author} {\bibfnamefont {M.}~\bibnamefont {{Richter}}},
  \bibinfo {author} {\bibfnamefont {D.}~\bibnamefont {{Rigopoulou}}}, \bibinfo
  {author} {\bibfnamefont {D.}~\bibnamefont {{Sanders}}}, \bibinfo {author}
  {\bibfnamefont {R.}~\bibnamefont {{Sankrit}}}, \bibinfo {author}
  {\bibfnamefont {G.}~\bibnamefont {{Savini}}}, \bibinfo {author}
  {\bibfnamefont {J.-D.}\ \bibnamefont {{Smith}}}, \ and\ \bibinfo {author}
  {\bibfnamefont {S.}~\bibnamefont {{Stierwalt}}},\ }\bibfield  {title}
  {\enquote {\bibinfo {title} {{Review: far-infrared instrumentation and
  technological development for the next decade}},}\ }\href {\doibase
  10.1117/1.JATIS.5.2.020901} {\bibfield  {journal} {\bibinfo  {journal}
  {Journal of Astronomical Telescopes, Instruments, and Systems}\ }\textbf
  {\bibinfo {volume} {5}},\ \bibinfo {eid} {020901} (\bibinfo {year}
  {2019})}\BibitemShut {NoStop}%
\bibitem [{\citenamefont {Hailey-Dunsheath}\ \emph {et~al.}(2021)\citenamefont
  {Hailey-Dunsheath}, \citenamefont {Janssen}, \citenamefont {Glenn},
  \citenamefont {Bradford}, \citenamefont {Perido}, \citenamefont {Redford},\
  and\ \citenamefont {Zmuidzinas}}]{Dunsheath2021}%
  \BibitemOpen
  \bibfield  {author} {\bibinfo {author} {\bibfnamefont {S.}~\bibnamefont
  {Hailey-Dunsheath}}, \bibinfo {author} {\bibfnamefont {R.~M.~J.}\
  \bibnamefont {Janssen}}, \bibinfo {author} {\bibfnamefont {J.}~\bibnamefont
  {Glenn}}, \bibinfo {author} {\bibfnamefont {C.~M.}\ \bibnamefont {Bradford}},
  \bibinfo {author} {\bibfnamefont {J.}~\bibnamefont {Perido}}, \bibinfo
  {author} {\bibfnamefont {J.}~\bibnamefont {Redford}}, \ and\ \bibinfo
  {author} {\bibfnamefont {J.}~\bibnamefont {Zmuidzinas}},\ }\bibfield  {title}
  {\enquote {\bibinfo {title} {{Kinetic inductance detectors for the Origins
  Space Telescope}},}\ }\href {\doibase 10.1117/1.JATIS.7.1.011015} {\bibfield
  {journal} {\bibinfo  {journal} {Journal of Astronomical Telescopes,
  Instruments, and Systems}\ }\textbf {\bibinfo {volume} {7}},\ \bibinfo
  {pages} {1 -- 17} (\bibinfo {year} {2021})}\BibitemShut {NoStop}%
\bibitem [{\citenamefont {Irwin}\ and\ \citenamefont
  {Hilton}(2005)}]{Irwin2005}%
  \BibitemOpen
  \bibfield  {author} {\bibinfo {author} {\bibfnamefont {K.}~\bibnamefont
  {Irwin}}\ and\ \bibinfo {author} {\bibfnamefont {G.}~\bibnamefont {Hilton}},\
  }\enquote {\bibinfo {title} {Transition-edge sensors},}\ in\ \href {\doibase
  10.1007/10933596_3} {\emph {\bibinfo {booktitle} {Cryogenic Particle
  Detection}}},\ \bibinfo {editor} {edited by\ \bibinfo {editor} {\bibfnamefont
  {C.}~\bibnamefont {Enss}}}\ (\bibinfo  {publisher} {Springer Berlin
  Heidelberg},\ \bibinfo {address} {Berlin, Heidelberg},\ \bibinfo {year}
  {2005})\ pp.\ \bibinfo {pages} {63--150}\BibitemShut {NoStop}%
\bibitem [{\citenamefont {{Audley}}\ \emph {et~al.}(2016)\citenamefont
  {{Audley}}, \citenamefont {{de Lange}}, \citenamefont {{Gao}}, \citenamefont
  {{Khosropanah}}, \citenamefont {{Hijmering}}, \citenamefont {{Ridder}},
  \citenamefont {{Mauskopf}}, \citenamefont {{Morozov}}, \citenamefont
  {{Trappe}},\ and\ \citenamefont {{Doherty}}}]{Audley2016}%
  \BibitemOpen
  \bibfield  {author} {\bibinfo {author} {\bibfnamefont {M.~D.}\ \bibnamefont
  {{Audley}}}, \bibinfo {author} {\bibfnamefont {G.}~\bibnamefont {{de
  Lange}}}, \bibinfo {author} {\bibfnamefont {J.-R.}\ \bibnamefont {{Gao}}},
  \bibinfo {author} {\bibfnamefont {P.}~\bibnamefont {{Khosropanah}}}, \bibinfo
  {author} {\bibfnamefont {R.}~\bibnamefont {{Hijmering}}}, \bibinfo {author}
  {\bibfnamefont {M.}~\bibnamefont {{Ridder}}}, \bibinfo {author}
  {\bibfnamefont {P.~D.}\ \bibnamefont {{Mauskopf}}}, \bibinfo {author}
  {\bibfnamefont {D.}~\bibnamefont {{Morozov}}}, \bibinfo {author}
  {\bibfnamefont {N.~A.}\ \bibnamefont {{Trappe}}}, \ and\ \bibinfo {author}
  {\bibfnamefont {S.}~\bibnamefont {{Doherty}}},\ }\bibfield  {title} {\enquote
  {\bibinfo {title} {{Optical performance of an ultra-sensitive horn-coupled
  transition-edge-sensor bolometer with hemispherical backshort in the far
  infrared}},}\ }\href {\doibase 10.1063/1.4945302} {\bibfield  {journal}
  {\bibinfo  {journal} {Review of Scientific Instruments}\ }\textbf {\bibinfo
  {volume} {87}},\ \bibinfo {eid} {043103} (\bibinfo {year} {2016})},\ \Eprint
  {http://arxiv.org/abs/1603.07944} {arXiv:1603.07944 [astro-ph.IM]}
  \BibitemShut {NoStop}%
\bibitem [{\citenamefont {{Khosropanah}}\ \emph {et~al.}(2016)\citenamefont
  {{Khosropanah}}, \citenamefont {{Suzuki}}, \citenamefont {{Ridder}},
  \citenamefont {{Hijmering}}, \citenamefont {{Akamatsu}}, \citenamefont
  {{Gottardi}}, \citenamefont {{van der Kuur}}, \citenamefont {{Gao}},\ and\
  \citenamefont {{Jackson}}}]{Pourya2016}%
  \BibitemOpen
  \bibfield  {author} {\bibinfo {author} {\bibfnamefont {P.}~\bibnamefont
  {{Khosropanah}}}, \bibinfo {author} {\bibfnamefont {T.}~\bibnamefont
  {{Suzuki}}}, \bibinfo {author} {\bibfnamefont {M.~L.}\ \bibnamefont
  {{Ridder}}}, \bibinfo {author} {\bibfnamefont {R.~A.}\ \bibnamefont
  {{Hijmering}}}, \bibinfo {author} {\bibfnamefont {H.}~\bibnamefont
  {{Akamatsu}}}, \bibinfo {author} {\bibfnamefont {L.}~\bibnamefont
  {{Gottardi}}}, \bibinfo {author} {\bibfnamefont {J.}~\bibnamefont {{van der
  Kuur}}}, \bibinfo {author} {\bibfnamefont {J.~R.}\ \bibnamefont {{Gao}}}, \
  and\ \bibinfo {author} {\bibfnamefont {B.~D.}\ \bibnamefont {{Jackson}}},\
  }\bibfield  {title} {\enquote {\bibinfo {title} {{Ultra-low noise TES
  bolometer arrays for SAFARI instrument on SPICA}},}\ }in\ \href {\doibase
  10.1117/12.2233472} {\emph {\bibinfo {booktitle} {Millimeter, Submillimeter,
  and Far-Infrared Detectors and Instrumentation for Astronomy VIII}}},\
  \bibinfo {series} {Society of Photo-Optical Instrumentation Engineers (SPIE)
  Conference Series}, Vol.\ \bibinfo {volume} {9914},\ \bibinfo {editor}
  {edited by\ \bibinfo {editor} {\bibfnamefont {W.~S.}\ \bibnamefont
  {{Holland}}}\ and\ \bibinfo {editor} {\bibfnamefont {J.}~\bibnamefont
  {{Zmuidzinas}}}}\ (\bibinfo {year} {2016})\ p.\ \bibinfo {pages}
  {99140B}\BibitemShut {NoStop}%
\bibitem [{\citenamefont {Williams}\ \emph {et~al.}(2020)\citenamefont
  {Williams}, \citenamefont {Withington}, \citenamefont {Goldie}, \citenamefont
  {Thomas}, \citenamefont {Ade},\ and\ \citenamefont
  {Sudiwala}}]{Williams2020}%
  \BibitemOpen
  \bibfield  {author} {\bibinfo {author} {\bibfnamefont {E.~A.}\ \bibnamefont
  {Williams}}, \bibinfo {author} {\bibfnamefont {S.}~\bibnamefont
  {Withington}}, \bibinfo {author} {\bibfnamefont {D.~J.}\ \bibnamefont
  {Goldie}}, \bibinfo {author} {\bibfnamefont {C.~N.}\ \bibnamefont {Thomas}},
  \bibinfo {author} {\bibfnamefont {P.~A.~R.}\ \bibnamefont {Ade}}, \ and\
  \bibinfo {author} {\bibfnamefont {R.}~\bibnamefont {Sudiwala}},\ }\bibfield
  {title} {\enquote {\bibinfo {title} {Characterizing the optical response of
  ultra-low-noise far-infrared 60–110 $\mu$m transition edge sensors},}\
  }\href {\doibase 10.1063/5.0025900} {\bibfield  {journal} {\bibinfo
  {journal} {Review of Scientific Instruments}\ }\textbf {\bibinfo {volume}
  {91}},\ \bibinfo {pages} {123104} (\bibinfo {year} {2020})},\ \Eprint
  {http://arxiv.org/abs/https://doi.org/10.1063/5.0025900}
  {https://doi.org/10.1063/5.0025900} \BibitemShut {NoStop}%
\bibitem [{\citenamefont {{Nagler}}, \citenamefont {{Sadleir}},\ and\
  \citenamefont {{Wollack}}(2020)}]{Nagler2020}%
  \BibitemOpen
  \bibfield  {author} {\bibinfo {author} {\bibfnamefont {P.~C.}\ \bibnamefont
  {{Nagler}}}, \bibinfo {author} {\bibfnamefont {J.~E.}\ \bibnamefont
  {{Sadleir}}}, \ and\ \bibinfo {author} {\bibfnamefont {E.~J.}\ \bibnamefont
  {{Wollack}}},\ }\bibfield  {title} {\enquote {\bibinfo {title}
  {{Demonstration of ultra-low noise equivalent power using a longitudinal
  proximity effect transition-edge sensor}},}\ }\href@noop {} {\bibfield
  {journal} {\bibinfo  {journal} {arXiv e-prints}\ ,\ \bibinfo {eid}
  {arXiv:2012.06543}} (\bibinfo {year} {2020})},\ \Eprint
  {http://arxiv.org/abs/2012.06543} {arXiv:2012.06543 [astro-ph.IM]}
  \BibitemShut {NoStop}%
\bibitem [{\citenamefont {Echternach}\ \emph {et~al.}(2013)\citenamefont
  {Echternach}, \citenamefont {Stone}, \citenamefont {Bradford}, \citenamefont
  {Day}, \citenamefont {Wilson}, \citenamefont {Megerian}, \citenamefont
  {Llombart},\ and\ \citenamefont {Bueno}}]{Echternach2013}%
  \BibitemOpen
  \bibfield  {author} {\bibinfo {author} {\bibfnamefont {P.~M.}\ \bibnamefont
  {Echternach}}, \bibinfo {author} {\bibfnamefont {K.~J.}\ \bibnamefont
  {Stone}}, \bibinfo {author} {\bibfnamefont {C.~M.}\ \bibnamefont {Bradford}},
  \bibinfo {author} {\bibfnamefont {P.~K.}\ \bibnamefont {Day}}, \bibinfo
  {author} {\bibfnamefont {D.~W.}\ \bibnamefont {Wilson}}, \bibinfo {author}
  {\bibfnamefont {K.~G.}\ \bibnamefont {Megerian}}, \bibinfo {author}
  {\bibfnamefont {N.}~\bibnamefont {Llombart}}, \ and\ \bibinfo {author}
  {\bibfnamefont {J.}~\bibnamefont {Bueno}},\ }\bibfield  {title} {\enquote
  {\bibinfo {title} {Photon shot noise limited detection of terahertz radiation
  using a quantum capacitance detector},}\ }\href {\doibase
  http://dx.doi.org/10.1063/1.4817585} {\bibfield  {journal} {\bibinfo
  {journal} {Appl. Phys. Lett.}\ }\textbf {\bibinfo {volume} {103}},\ \bibinfo
  {eid} {053510} (\bibinfo {year} {2013})}\BibitemShut {NoStop}%
\bibitem [{\citenamefont {Echternach}, \citenamefont {Beyer},\ and\
  \citenamefont {Bradford}(2021)}]{Echternach2021}%
  \BibitemOpen
  \bibfield  {author} {\bibinfo {author} {\bibfnamefont {P.~M.}\ \bibnamefont
  {Echternach}}, \bibinfo {author} {\bibfnamefont {A.~D.}\ \bibnamefont
  {Beyer}}, \ and\ \bibinfo {author} {\bibfnamefont {C.~M.}\ \bibnamefont
  {Bradford}},\ }\bibfield  {title} {\enquote {\bibinfo {title} {{Large array
  of low-frequency readout quantum capacitance detectors}},}\ }\href {\doibase
  10.1117/1.JATIS.7.1.011003} {\bibfield  {journal} {\bibinfo  {journal}
  {Journal of Astronomical Telescopes, Instruments, and Systems}\ }\textbf
  {\bibinfo {volume} {7}},\ \bibinfo {pages} {1 -- 8} (\bibinfo {year}
  {2021})}\BibitemShut {NoStop}%
\bibitem [{\citenamefont {Day}\ \emph {et~al.}(2003)\citenamefont {Day},
  \citenamefont {LeDuc}, \citenamefont {Mazin}, \citenamefont {Vayonakis},\
  and\ \citenamefont {Zmuidzinas}}]{Day2003}%
  \BibitemOpen
  \bibfield  {author} {\bibinfo {author} {\bibfnamefont {P.~K.}\ \bibnamefont
  {Day}}, \bibinfo {author} {\bibfnamefont {H.~G.}\ \bibnamefont {LeDuc}},
  \bibinfo {author} {\bibfnamefont {B.~A.}\ \bibnamefont {Mazin}}, \bibinfo
  {author} {\bibfnamefont {A.}~\bibnamefont {Vayonakis}}, \ and\ \bibinfo
  {author} {\bibfnamefont {J.}~\bibnamefont {Zmuidzinas}},\ }\bibfield  {title}
  {\enquote {\bibinfo {title} {A broadband superconducting detector suitable
  for use in large arrays},}\ }\href@noop {} {\bibfield  {journal} {\bibinfo
  {journal} {Nature}\ }\textbf {\bibinfo {volume} {425}},\ \bibinfo {pages}
  {817} (\bibinfo {year} {2003})}\BibitemShut {NoStop}%
\bibitem [{\citenamefont {{Baselmans}}\ \emph {et~al.}(2017)\citenamefont
  {{Baselmans}}, \citenamefont {{Bueno}}, \citenamefont {{Yates}},
  \citenamefont {{Yurduseven}}, \citenamefont {{Llombart}}, \citenamefont
  {{Karatsu}}, \citenamefont {{Baryshev}}, \citenamefont {{Ferrari}},
  \citenamefont {{Endo}}, \citenamefont {{Thoen}}, \citenamefont {{de Visser}},
  \citenamefont {{Janssen}}, \citenamefont {{Murugesan}}, \citenamefont
  {{Driessen}}, \citenamefont {{Coiffard}}, \citenamefont {{Martin-Pintado}},
  \citenamefont {{Hargrave}},\ and\ \citenamefont {{Griffin}}}]{Baselmans2017}%
  \BibitemOpen
  \bibfield  {author} {\bibinfo {author} {\bibfnamefont {J.~J.~A.}\
  \bibnamefont {{Baselmans}}}, \bibinfo {author} {\bibfnamefont
  {J.}~\bibnamefont {{Bueno}}}, \bibinfo {author} {\bibfnamefont {S.~J.~C.}\
  \bibnamefont {{Yates}}}, \bibinfo {author} {\bibfnamefont {O.}~\bibnamefont
  {{Yurduseven}}}, \bibinfo {author} {\bibfnamefont {N.}~\bibnamefont
  {{Llombart}}}, \bibinfo {author} {\bibfnamefont {K.}~\bibnamefont
  {{Karatsu}}}, \bibinfo {author} {\bibfnamefont {A.~M.}\ \bibnamefont
  {{Baryshev}}}, \bibinfo {author} {\bibfnamefont {L.}~\bibnamefont
  {{Ferrari}}}, \bibinfo {author} {\bibfnamefont {A.}~\bibnamefont {{Endo}}},
  \bibinfo {author} {\bibfnamefont {D.~J.}\ \bibnamefont {{Thoen}}}, \bibinfo
  {author} {\bibfnamefont {P.~J.}\ \bibnamefont {{de Visser}}}, \bibinfo
  {author} {\bibfnamefont {R.~M.~J.}\ \bibnamefont {{Janssen}}}, \bibinfo
  {author} {\bibfnamefont {V.}~\bibnamefont {{Murugesan}}}, \bibinfo {author}
  {\bibfnamefont {E.~F.~C.}\ \bibnamefont {{Driessen}}}, \bibinfo {author}
  {\bibfnamefont {G.}~\bibnamefont {{Coiffard}}}, \bibinfo {author}
  {\bibfnamefont {J.}~\bibnamefont {{Martin-Pintado}}}, \bibinfo {author}
  {\bibfnamefont {P.}~\bibnamefont {{Hargrave}}}, \ and\ \bibinfo {author}
  {\bibfnamefont {M.}~\bibnamefont {{Griffin}}},\ }\bibfield  {title} {\enquote
  {\bibinfo {title} {{A kilo-pixel imaging system for future space based
  far-infrared observatories using microwave kinetic inductance detectors}},}\
  }\href {\doibase 10.1051/0004-6361/201629653} {\bibfield  {journal} {\bibinfo
   {journal} {Astronomy and Astrophysics}\ }\textbf {\bibinfo {volume} {601}},\
  \bibinfo {eid} {A89} (\bibinfo {year} {2017})},\ \Eprint
  {http://arxiv.org/abs/1609.01952} {arXiv:1609.01952 [astro-ph.IM]}
  \BibitemShut {NoStop}%
\bibitem [{\citenamefont {{Bueno}}\ \emph {et~al.}(2017)\citenamefont
  {{Bueno}}, \citenamefont {{Yurduseven}}, \citenamefont {{Yates}},
  \citenamefont {{Llombart}}, \citenamefont {{Murugesan}}, \citenamefont
  {{Thoen}}, \citenamefont {{Baryshev}}, \citenamefont {{Neto}},\ and\
  \citenamefont {{Baselmans}}}]{Bueno2017}%
  \BibitemOpen
  \bibfield  {author} {\bibinfo {author} {\bibfnamefont {J.}~\bibnamefont
  {{Bueno}}}, \bibinfo {author} {\bibfnamefont {O.}~\bibnamefont
  {{Yurduseven}}}, \bibinfo {author} {\bibfnamefont {S.~J.~C.}\ \bibnamefont
  {{Yates}}}, \bibinfo {author} {\bibfnamefont {N.}~\bibnamefont {{Llombart}}},
  \bibinfo {author} {\bibfnamefont {V.}~\bibnamefont {{Murugesan}}}, \bibinfo
  {author} {\bibfnamefont {D.~J.}\ \bibnamefont {{Thoen}}}, \bibinfo {author}
  {\bibfnamefont {A.~M.}\ \bibnamefont {{Baryshev}}}, \bibinfo {author}
  {\bibfnamefont {A.}~\bibnamefont {{Neto}}}, \ and\ \bibinfo {author}
  {\bibfnamefont {J.~J.~A.}\ \bibnamefont {{Baselmans}}},\ }\bibfield  {title}
  {\enquote {\bibinfo {title} {{Full characterisation of a background limited
  antenna coupled KID over an octave of bandwidth for THz radiation}},}\ }\href
  {\doibase 10.1063/1.4985060} {\bibfield  {journal} {\bibinfo  {journal}
  {Applied Physics Letters}\ }\textbf {\bibinfo {volume} {110}},\ \bibinfo
  {eid} {233503} (\bibinfo {year} {2017})}\BibitemShut {NoStop}%
\bibitem [{\citenamefont {{Bueno}}\ \emph {et~al.}(2018)\citenamefont
  {{Bueno}}, \citenamefont {{Murugesan}}, \citenamefont {{Karatsu}},
  \citenamefont {{Thoen}},\ and\ \citenamefont
  {{Baselmans}}}]{Bueno2017kpixel}%
  \BibitemOpen
  \bibfield  {author} {\bibinfo {author} {\bibfnamefont {J.}~\bibnamefont
  {{Bueno}}}, \bibinfo {author} {\bibfnamefont {V.}~\bibnamefont
  {{Murugesan}}}, \bibinfo {author} {\bibfnamefont {K.}~\bibnamefont
  {{Karatsu}}}, \bibinfo {author} {\bibfnamefont {D.~J.}\ \bibnamefont
  {{Thoen}}}, \ and\ \bibinfo {author} {\bibfnamefont {J.~J.~A.}\ \bibnamefont
  {{Baselmans}}},\ }\bibfield  {title} {\enquote {\bibinfo {title}
  {{Ultrasensitive Kilo-Pixel Imaging Array of Photon Noise-Limited Kinetic
  Inductance Detectors Over an Octave of Bandwidth for THz Astronomy}},}\
  }\href {\doibase 10.1007/s10909-018-1962-8} {\bibfield  {journal} {\bibinfo
  {journal} {Journal of Low Temperature Physics}\ }\textbf {\bibinfo {volume}
  {193}},\ \bibinfo {pages} {96--102} (\bibinfo {year} {2018})}\BibitemShut
  {NoStop}%
\bibitem [{\citenamefont {{McGeehan}}\ \emph {et~al.}(2018)\citenamefont
  {{McGeehan}}, \citenamefont {{Barry}}, \citenamefont {{Shirokoff}},
  \citenamefont {{Bradford}}, \citenamefont {{Che}}, \citenamefont {{Glenn}},
  \citenamefont {{Gordon}}, \citenamefont {{Hailey-Dunsheath}}, \citenamefont
  {{Hollister}}, \citenamefont {{Kov{\'a}cs}}, \citenamefont {{LeDuc}},
  \citenamefont {{Mauskopf}}, \citenamefont {{McKenney}}, \citenamefont
  {{Reck}}, \citenamefont {{Redford}}, \citenamefont {{Tucker}}, \citenamefont
  {{Turner}}, \citenamefont {{Walker}}, \citenamefont {{Wheeler}},\ and\
  \citenamefont {{Zmuidzinas}}}]{Geehan2018}%
  \BibitemOpen
  \bibfield  {author} {\bibinfo {author} {\bibfnamefont {R.}~\bibnamefont
  {{McGeehan}}}, \bibinfo {author} {\bibfnamefont {P.~S.}\ \bibnamefont
  {{Barry}}}, \bibinfo {author} {\bibfnamefont {E.}~\bibnamefont
  {{Shirokoff}}}, \bibinfo {author} {\bibfnamefont {C.~M.}\ \bibnamefont
  {{Bradford}}}, \bibinfo {author} {\bibfnamefont {G.}~\bibnamefont {{Che}}},
  \bibinfo {author} {\bibfnamefont {J.}~\bibnamefont {{Glenn}}}, \bibinfo
  {author} {\bibfnamefont {S.}~\bibnamefont {{Gordon}}}, \bibinfo {author}
  {\bibfnamefont {S.}~\bibnamefont {{Hailey-Dunsheath}}}, \bibinfo {author}
  {\bibfnamefont {M.}~\bibnamefont {{Hollister}}}, \bibinfo {author}
  {\bibfnamefont {A.}~\bibnamefont {{Kov{\'a}cs}}}, \bibinfo {author}
  {\bibfnamefont {H.~G.}\ \bibnamefont {{LeDuc}}}, \bibinfo {author}
  {\bibfnamefont {P.}~\bibnamefont {{Mauskopf}}}, \bibinfo {author}
  {\bibfnamefont {C.}~\bibnamefont {{McKenney}}}, \bibinfo {author}
  {\bibfnamefont {T.}~\bibnamefont {{Reck}}}, \bibinfo {author} {\bibfnamefont
  {J.}~\bibnamefont {{Redford}}}, \bibinfo {author} {\bibfnamefont
  {C.}~\bibnamefont {{Tucker}}}, \bibinfo {author} {\bibfnamefont
  {J.}~\bibnamefont {{Turner}}}, \bibinfo {author} {\bibfnamefont
  {S.}~\bibnamefont {{Walker}}}, \bibinfo {author} {\bibfnamefont
  {J.}~\bibnamefont {{Wheeler}}}, \ and\ \bibinfo {author} {\bibfnamefont
  {J.}~\bibnamefont {{Zmuidzinas}}},\ }\bibfield  {title} {\enquote {\bibinfo
  {title} {{Low-Temperature Noise Performance of SuperSpec and Other
  Developments on the Path to Deployment}},}\ }\href {\doibase
  10.1007/s10909-018-2061-6} {\bibfield  {journal} {\bibinfo  {journal}
  {Journal of Low Temperature Physics}\ }\textbf {\bibinfo {volume} {193}},\
  \bibinfo {pages} {1024--1032} (\bibinfo {year} {2018})}\BibitemShut {NoStop}%
\bibitem [{\citenamefont {{Karatsu}}\ \emph {et~al.}(2019)\citenamefont
  {{Karatsu}}, \citenamefont {{Endo}}, \citenamefont {{Bueno}}, \citenamefont
  {{de Visser}}, \citenamefont {{Barends}}, \citenamefont {{Thoen}},
  \citenamefont {{Murugesan}}, \citenamefont {{Tomita}},\ and\ \citenamefont
  {{Baselmans}}}]{Karatsu2019}%
  \BibitemOpen
  \bibfield  {author} {\bibinfo {author} {\bibfnamefont {K.}~\bibnamefont
  {{Karatsu}}}, \bibinfo {author} {\bibfnamefont {A.}~\bibnamefont {{Endo}}},
  \bibinfo {author} {\bibfnamefont {J.}~\bibnamefont {{Bueno}}}, \bibinfo
  {author} {\bibfnamefont {P.~J.}\ \bibnamefont {{de Visser}}}, \bibinfo
  {author} {\bibfnamefont {R.}~\bibnamefont {{Barends}}}, \bibinfo {author}
  {\bibfnamefont {D.~J.}\ \bibnamefont {{Thoen}}}, \bibinfo {author}
  {\bibfnamefont {V.}~\bibnamefont {{Murugesan}}}, \bibinfo {author}
  {\bibfnamefont {N.}~\bibnamefont {{Tomita}}}, \ and\ \bibinfo {author}
  {\bibfnamefont {J.~J.~A.}\ \bibnamefont {{Baselmans}}},\ }\bibfield  {title}
  {\enquote {\bibinfo {title} {{Mitigation of cosmic ray effect on microwave
  kinetic inductance detector arrays}},}\ }\href {\doibase 10.1063/1.5052419}
  {\bibfield  {journal} {\bibinfo  {journal} {Applied Physics Letters}\
  }\textbf {\bibinfo {volume} {114}},\ \bibinfo {eid} {032601} (\bibinfo {year}
  {2019})},\ \Eprint {http://arxiv.org/abs/1901.02387} {arXiv:1901.02387
  [astro-ph.IM]} \BibitemShut {NoStop}%
\bibitem [{\citenamefont {{Karatsu}}\ \emph {et~al.}(2016)\citenamefont
  {{Karatsu}}, \citenamefont {{Dominjon}}, \citenamefont {{Fujino}},
  \citenamefont {{Funaki}}, \citenamefont {{Hazumi}}, \citenamefont {{Irie}},
  \citenamefont {{Ishino}}, \citenamefont {{Kida}}, \citenamefont
  {{Matsumura}}, \citenamefont {{Mizukami}}, \citenamefont {{Naruse}},
  \citenamefont {{Nitta}}, \citenamefont {{Noguchi}}, \citenamefont {{Oka}},
  \citenamefont {{Sekiguchi}}, \citenamefont {{Sekimoto}}, \citenamefont
  {{Sekine}}, \citenamefont {{Shu}}, \citenamefont {{Yamada}},\ and\
  \citenamefont {{Yamashita}}}]{Karatsu2016}%
  \BibitemOpen
  \bibfield  {author} {\bibinfo {author} {\bibfnamefont {K.}~\bibnamefont
  {{Karatsu}}}, \bibinfo {author} {\bibfnamefont {A.}~\bibnamefont
  {{Dominjon}}}, \bibinfo {author} {\bibfnamefont {T.}~\bibnamefont
  {{Fujino}}}, \bibinfo {author} {\bibfnamefont {T.}~\bibnamefont {{Funaki}}},
  \bibinfo {author} {\bibfnamefont {M.}~\bibnamefont {{Hazumi}}}, \bibinfo
  {author} {\bibfnamefont {F.}~\bibnamefont {{Irie}}}, \bibinfo {author}
  {\bibfnamefont {H.}~\bibnamefont {{Ishino}}}, \bibinfo {author}
  {\bibfnamefont {Y.}~\bibnamefont {{Kida}}}, \bibinfo {author} {\bibfnamefont
  {T.}~\bibnamefont {{Matsumura}}}, \bibinfo {author} {\bibfnamefont
  {K.}~\bibnamefont {{Mizukami}}}, \bibinfo {author} {\bibfnamefont
  {M.}~\bibnamefont {{Naruse}}}, \bibinfo {author} {\bibfnamefont
  {T.}~\bibnamefont {{Nitta}}}, \bibinfo {author} {\bibfnamefont
  {T.}~\bibnamefont {{Noguchi}}}, \bibinfo {author} {\bibfnamefont
  {N.}~\bibnamefont {{Oka}}}, \bibinfo {author} {\bibfnamefont
  {S.}~\bibnamefont {{Sekiguchi}}}, \bibinfo {author} {\bibfnamefont
  {Y.}~\bibnamefont {{Sekimoto}}}, \bibinfo {author} {\bibfnamefont
  {M.}~\bibnamefont {{Sekine}}}, \bibinfo {author} {\bibfnamefont
  {S.}~\bibnamefont {{Shu}}}, \bibinfo {author} {\bibfnamefont
  {Y.}~\bibnamefont {{Yamada}}}, \ and\ \bibinfo {author} {\bibfnamefont
  {T.}~\bibnamefont {{Yamashita}}},\ }\bibfield  {title} {\enquote {\bibinfo
  {title} {{Radiation Tolerance of Aluminum Microwave Kinetic Inductance
  Detector}},}\ }\href {\doibase 10.1007/s10909-016-1523-y} {\bibfield
  {journal} {\bibinfo  {journal} {Journal of Low Temperature Physics}\ }\textbf
  {\bibinfo {volume} {184}},\ \bibinfo {pages} {540--546} (\bibinfo {year}
  {2016})}\BibitemShut {NoStop}%
\bibitem [{\citenamefont {de~Visser}\ \emph {et~al.}(2014)\citenamefont
  {de~Visser}, \citenamefont {Baselmans}, \citenamefont {Bueno}, \citenamefont
  {Llombart},\ and\ \citenamefont {Klapwijk}}]{Visser2014}%
  \BibitemOpen
  \bibfield  {author} {\bibinfo {author} {\bibfnamefont {P.~J.}\ \bibnamefont
  {de~Visser}}, \bibinfo {author} {\bibfnamefont {J.~J.~A.}\ \bibnamefont
  {Baselmans}}, \bibinfo {author} {\bibfnamefont {J.}~\bibnamefont {Bueno}},
  \bibinfo {author} {\bibfnamefont {N.}~\bibnamefont {Llombart}}, \ and\
  \bibinfo {author} {\bibfnamefont {T.~M.}\ \bibnamefont {Klapwijk}},\
  }\bibfield  {title} {\enquote {\bibinfo {title} {Fluctuations in the electron
  system of a superconductor exposed to a photon flux},}\ }\href {\doibase
  10.1038/ncomms4130} {\bibfield  {journal} {\bibinfo  {journal} {Nat.
  Commun.}\ }\textbf {\bibinfo {volume} {5}},\ \bibinfo {pages} {3130}
  (\bibinfo {year} {2014})}\BibitemShut {NoStop}%
\bibitem [{\citenamefont {{Flanigan}}\ \emph {et~al.}(2016)\citenamefont
  {{Flanigan}}, \citenamefont {{McCarrick}}, \citenamefont {{Jones}},
  \citenamefont {{Johnson}}, \citenamefont {{Abitbol}}, \citenamefont {{Ade}},
  \citenamefont {{Araujo}}, \citenamefont {{Bradford}}, \citenamefont
  {{Cantor}}, \citenamefont {{Che}}, \citenamefont {{Day}}, \citenamefont
  {{Doyle}}, \citenamefont {{Kjellstrand}}, \citenamefont {{Leduc}},
  \citenamefont {{Limon}}, \citenamefont {{Luu}}, \citenamefont {{Mauskopf}},
  \citenamefont {{Miller}}, \citenamefont {{Mroczkowski}}, \citenamefont
  {{Tucker}},\ and\ \citenamefont {{Zmuidzinas}}}]{Flanigan2017}%
  \BibitemOpen
  \bibfield  {author} {\bibinfo {author} {\bibfnamefont {D.}~\bibnamefont
  {{Flanigan}}}, \bibinfo {author} {\bibfnamefont {H.}~\bibnamefont
  {{McCarrick}}}, \bibinfo {author} {\bibfnamefont {G.}~\bibnamefont
  {{Jones}}}, \bibinfo {author} {\bibfnamefont {B.~R.}\ \bibnamefont
  {{Johnson}}}, \bibinfo {author} {\bibfnamefont {M.~H.}\ \bibnamefont
  {{Abitbol}}}, \bibinfo {author} {\bibfnamefont {P.}~\bibnamefont {{Ade}}},
  \bibinfo {author} {\bibfnamefont {D.}~\bibnamefont {{Araujo}}}, \bibinfo
  {author} {\bibfnamefont {K.}~\bibnamefont {{Bradford}}}, \bibinfo {author}
  {\bibfnamefont {R.}~\bibnamefont {{Cantor}}}, \bibinfo {author}
  {\bibfnamefont {G.}~\bibnamefont {{Che}}}, \bibinfo {author} {\bibfnamefont
  {P.}~\bibnamefont {{Day}}}, \bibinfo {author} {\bibfnamefont
  {S.}~\bibnamefont {{Doyle}}}, \bibinfo {author} {\bibfnamefont {C.~B.}\
  \bibnamefont {{Kjellstrand}}}, \bibinfo {author} {\bibfnamefont
  {H.}~\bibnamefont {{Leduc}}}, \bibinfo {author} {\bibfnamefont
  {M.}~\bibnamefont {{Limon}}}, \bibinfo {author} {\bibfnamefont
  {V.}~\bibnamefont {{Luu}}}, \bibinfo {author} {\bibfnamefont
  {P.}~\bibnamefont {{Mauskopf}}}, \bibinfo {author} {\bibfnamefont
  {A.}~\bibnamefont {{Miller}}}, \bibinfo {author} {\bibfnamefont
  {T.}~\bibnamefont {{Mroczkowski}}}, \bibinfo {author} {\bibfnamefont
  {C.}~\bibnamefont {{Tucker}}}, \ and\ \bibinfo {author} {\bibfnamefont
  {J.}~\bibnamefont {{Zmuidzinas}}},\ }\bibfield  {title} {\enquote {\bibinfo
  {title} {{Photon noise from chaotic and coherent millimeter-wave sources
  measured with horn-coupled, aluminum lumped-element kinetic inductance
  detectors}},}\ }\href {\doibase 10.1063/1.4942804} {\bibfield  {journal}
  {\bibinfo  {journal} {Applied Physics Letters}\ }\textbf {\bibinfo {volume}
  {108}},\ \bibinfo {eid} {083504} (\bibinfo {year} {2016})},\ \Eprint
  {http://arxiv.org/abs/1510.06609} {arXiv:1510.06609 [astro-ph.IM]}
  \BibitemShut {NoStop}%
\bibitem [{\citenamefont {{Endo}}\ \emph {et~al.}(2020)\citenamefont {{Endo}},
  \citenamefont {{Laguna}}, \citenamefont {{H{\"a}hnle}}, \citenamefont
  {{Karatsu}}, \citenamefont {{Thoen}}, \citenamefont {{Murugesan}},\ and\
  \citenamefont {{Baselmans}}}]{Endo2020SONNET}%
  \BibitemOpen
  \bibfield  {author} {\bibinfo {author} {\bibfnamefont {A.}~\bibnamefont
  {{Endo}}}, \bibinfo {author} {\bibfnamefont {A.~P.}\ \bibnamefont
  {{Laguna}}}, \bibinfo {author} {\bibfnamefont {S.}~\bibnamefont
  {{H{\"a}hnle}}}, \bibinfo {author} {\bibfnamefont {K.}~\bibnamefont
  {{Karatsu}}}, \bibinfo {author} {\bibfnamefont {D.~J.}\ \bibnamefont
  {{Thoen}}}, \bibinfo {author} {\bibfnamefont {V.}~\bibnamefont
  {{Murugesan}}}, \ and\ \bibinfo {author} {\bibfnamefont {J.~J.~A.}\
  \bibnamefont {{Baselmans}}},\ }\bibfield  {title} {\enquote {\bibinfo {title}
  {{Simulating the radiation loss of superconducting submillimeter wave filters
  and transmission lines using Sonnet EM}},}\ }\href@noop {} {\bibfield
  {journal} {\bibinfo  {journal} {arXiv e-prints}\ ,\ \bibinfo {eid}
  {arXiv:2012.07251}} (\bibinfo {year} {2020})},\ \Eprint
  {http://arxiv.org/abs/2012.07251} {arXiv:2012.07251 [astro-ph.IM]}
  \BibitemShut {NoStop}%
\bibitem [{\citenamefont {Andreev}(1964)}]{Andreev1964}%
  \BibitemOpen
  \bibfield  {author} {\bibinfo {author} {\bibfnamefont {A.}~\bibnamefont
  {Andreev}},\ }\bibfield  {title} {\enquote {\bibinfo {title} {Thermal
  conductivity of the intermediate state of superconductors},}\ }\href@noop {}
  {\bibfield  {journal} {\bibinfo  {journal} {Zh. Eksperim. i Teor. Fiz.}\
  }\textbf {\bibinfo {volume} {46}} (\bibinfo {year} {1964})}\BibitemShut
  {NoStop}%
\bibitem [{\citenamefont {Kozorezov}\ \emph {et~al.}(2000)\citenamefont
  {Kozorezov}, \citenamefont {Volkov}, \citenamefont {Wigmore}, \citenamefont
  {Peacock}, \citenamefont {Poelaert},\ and\ \citenamefont {den
  Hartog}}]{Kozorezov2000}%
  \BibitemOpen
  \bibfield  {author} {\bibinfo {author} {\bibfnamefont {A.~G.}\ \bibnamefont
  {Kozorezov}}, \bibinfo {author} {\bibfnamefont {A.~F.}\ \bibnamefont
  {Volkov}}, \bibinfo {author} {\bibfnamefont {J.~K.}\ \bibnamefont {Wigmore}},
  \bibinfo {author} {\bibfnamefont {A.}~\bibnamefont {Peacock}}, \bibinfo
  {author} {\bibfnamefont {A.}~\bibnamefont {Poelaert}}, \ and\ \bibinfo
  {author} {\bibfnamefont {R.}~\bibnamefont {den Hartog}},\ }\bibfield  {title}
  {\enquote {\bibinfo {title} {Quasiparticle-phonon downconversion in
  nonequilibrium superconductors},}\ }\href@noop {} {\bibfield  {journal}
  {\bibinfo  {journal} {Phys. Rev. B}\ }\textbf {\bibinfo {volume} {61}},\
  \bibinfo {pages} {11807} (\bibinfo {year} {2000})}\BibitemShut {NoStop}%
\bibitem [{\citenamefont {Mattis}\ and\ \citenamefont
  {Bardeen}(1958)}]{Mattis1958}%
  \BibitemOpen
  \bibfield  {author} {\bibinfo {author} {\bibfnamefont {D.~C.}\ \bibnamefont
  {Mattis}}\ and\ \bibinfo {author} {\bibfnamefont {J.}~\bibnamefont
  {Bardeen}},\ }\bibfield  {title} {\enquote {\bibinfo {title} {Theory of the
  anomalous skin effect in normal and superconducting metals},}\ }\href@noop {}
  {\bibfield  {journal} {\bibinfo  {journal} {Phys. Rev.}\ }\textbf {\bibinfo
  {volume} {111}},\ \bibinfo {pages} {412} (\bibinfo {year}
  {1958})}\BibitemShut {NoStop}%
\bibitem [{\citenamefont {{Fyhrie}}\ \emph {et~al.}(2016)\citenamefont
  {{Fyhrie}}, \citenamefont {{McKenney}}, \citenamefont {{Glenn}},
  \citenamefont {{LeDuc}}, \citenamefont {{Gao}}, \citenamefont {{Day}},\ and\
  \citenamefont {{Zmuidzinas}}}]{Fyhrie2016}%
  \BibitemOpen
  \bibfield  {author} {\bibinfo {author} {\bibfnamefont {A.}~\bibnamefont
  {{Fyhrie}}}, \bibinfo {author} {\bibfnamefont {C.}~\bibnamefont
  {{McKenney}}}, \bibinfo {author} {\bibfnamefont {J.}~\bibnamefont {{Glenn}}},
  \bibinfo {author} {\bibfnamefont {H.~G.}\ \bibnamefont {{LeDuc}}}, \bibinfo
  {author} {\bibfnamefont {J.}~\bibnamefont {{Gao}}}, \bibinfo {author}
  {\bibfnamefont {P.}~\bibnamefont {{Day}}}, \ and\ \bibinfo {author}
  {\bibfnamefont {J.}~\bibnamefont {{Zmuidzinas}}},\ }\bibfield  {title}
  {\enquote {\bibinfo {title} {{Responsivity boosting in FIR TiN LEKIDs using
  phonon recycling: simulations and array design}},}\ }in\ \href {\doibase
  10.1117/12.2231476} {\emph {\bibinfo {booktitle} {Millimeter, Submillimeter,
  and Far-Infrared Detectors and Instrumentation for Astronomy VIII}}},\
  \bibinfo {series} {Society of Photo-Optical Instrumentation Engineers (SPIE)
  Conference Series}, Vol.\ \bibinfo {volume} {9914},\ \bibinfo {editor}
  {edited by\ \bibinfo {editor} {\bibfnamefont {W.~S.}\ \bibnamefont
  {{Holland}}}\ and\ \bibinfo {editor} {\bibfnamefont {J.}~\bibnamefont
  {{Zmuidzinas}}}}\ (\bibinfo {year} {2016})\ p.\ \bibinfo {pages}
  {99142B}\BibitemShut {NoStop}%
\bibitem [{\citenamefont {Kaplan}\ \emph {et~al.}(1976)\citenamefont {Kaplan},
  \citenamefont {Chi}, \citenamefont {Langenberg}, \citenamefont {Chang},
  \citenamefont {Jafarey},\ and\ \citenamefont {Scalapino}}]{Kaplan1976}%
  \BibitemOpen
  \bibfield  {author} {\bibinfo {author} {\bibfnamefont {S.~B.}\ \bibnamefont
  {Kaplan}}, \bibinfo {author} {\bibfnamefont {C.~C.}\ \bibnamefont {Chi}},
  \bibinfo {author} {\bibfnamefont {D.~N.}\ \bibnamefont {Langenberg}},
  \bibinfo {author} {\bibfnamefont {J.}~\bibnamefont {Chang}}, \bibinfo
  {author} {\bibfnamefont {S.}~\bibnamefont {Jafarey}}, \ and\ \bibinfo
  {author} {\bibfnamefont {D.~J.}\ \bibnamefont {Scalapino}},\ }\bibfield
  {title} {\enquote {\bibinfo {title} {Quasiparticle and phonon lifetimes in
  superconductors},}\ }\href@noop {} {\bibfield  {journal} {\bibinfo  {journal}
  {Phys. Rev. B}\ }\textbf {\bibinfo {volume} {14}},\ \bibinfo {pages}
  {4854--4873} (\bibinfo {year} {1976})}\BibitemShut {NoStop}%
\bibitem [{\citenamefont {Guruswamy}, \citenamefont {Goldie},\ and\
  \citenamefont {Withington}(2014)}]{Guruswamy2014}%
  \BibitemOpen
  \bibfield  {author} {\bibinfo {author} {\bibfnamefont {T.}~\bibnamefont
  {Guruswamy}}, \bibinfo {author} {\bibfnamefont {D.~J.}\ \bibnamefont
  {Goldie}}, \ and\ \bibinfo {author} {\bibfnamefont {S.}~\bibnamefont
  {Withington}},\ }\bibfield  {title} {\enquote {\bibinfo {title}
  {Quasiparticle generation efficiency in superconducting thin films},}\ }\href
  {\doibase 10.1088/0953-2048/27/5/055012} {\bibfield  {journal} {\bibinfo
  {journal} {Supercond. Sci. Technol.}\ }\textbf {\bibinfo {volume} {27}},\
  \bibinfo {pages} {055012} (\bibinfo {year} {2014})}\BibitemShut {NoStop}%
\bibitem [{\citenamefont {{de Visser}}\ \emph {et~al.}(2021)\citenamefont {{de
  Visser}}, \citenamefont {{de Rooij}}, \citenamefont {{Murugesan}},
  \citenamefont {{Thoen}},\ and\ \citenamefont {{Baselmans}}}]{2021deVisser}%
  \BibitemOpen
  \bibfield  {author} {\bibinfo {author} {\bibfnamefont {P.~J.}\ \bibnamefont
  {{de Visser}}}, \bibinfo {author} {\bibfnamefont {S.~A.~H.}\ \bibnamefont
  {{de Rooij}}}, \bibinfo {author} {\bibfnamefont {V.}~\bibnamefont
  {{Murugesan}}}, \bibinfo {author} {\bibfnamefont {D.~J.}\ \bibnamefont
  {{Thoen}}}, \ and\ \bibinfo {author} {\bibfnamefont {J.~J.~A.}\ \bibnamefont
  {{Baselmans}}},\ }\bibfield  {title} {\enquote {\bibinfo {title}
  {{Phonon-Trapping-Enhanced Energy Resolution in Superconducting Single-Photon
  Detectors}},}\ }\href {\doibase 10.1103/PhysRevApplied.16.034051} {\bibfield
  {journal} {\bibinfo  {journal} {Physical Review Applied}\ }\textbf {\bibinfo
  {volume} {16}},\ \bibinfo {eid} {034051} (\bibinfo {year} {2021})},\ \Eprint
  {http://arxiv.org/abs/2103.06723} {arXiv:2103.06723 [astro-ph.IM]}
  \BibitemShut {NoStop}%
\bibitem [{\citenamefont {Noroozian}\ \emph {et~al.}(2009)\citenamefont
  {Noroozian}, \citenamefont {Gao}, \citenamefont {Zmuidzinas}, \citenamefont
  {LeDuc},\ and\ \citenamefont {Mazin}}]{Noorozian2009}%
  \BibitemOpen
  \bibfield  {author} {\bibinfo {author} {\bibfnamefont {O.}~\bibnamefont
  {Noroozian}}, \bibinfo {author} {\bibfnamefont {J.}~\bibnamefont {Gao}},
  \bibinfo {author} {\bibfnamefont {J.}~\bibnamefont {Zmuidzinas}}, \bibinfo
  {author} {\bibfnamefont {H.~G.}\ \bibnamefont {LeDuc}}, \ and\ \bibinfo
  {author} {\bibfnamefont {B.~A.}\ \bibnamefont {Mazin}},\ }\bibfield  {title}
  {\enquote {\bibinfo {title} {Two-level system noise reduction for microwave
  kinetic inductance detectors},}\ }\href {\doibase
  http://dx.doi.org/10.1063/1.3292302} {\bibfield  {journal} {\bibinfo
  {journal} {AIP Conference Proceedings}\ }\textbf {\bibinfo {volume} {1185}},\
  \bibinfo {pages} {148--151} (\bibinfo {year} {2009})}\BibitemShut {NoStop}%
\bibitem [{\citenamefont {Gao}\ \emph {et~al.}(2008)\citenamefont {Gao},
  \citenamefont {Daal}, \citenamefont {Martinis}, \citenamefont {Vayonakis},
  \citenamefont {Zmuidzinas}, \citenamefont {Sadoulet}, \citenamefont {Mazin},
  \citenamefont {Day},\ and\ \citenamefont {Leduc}}]{Gao2008b}%
  \BibitemOpen
  \bibfield  {author} {\bibinfo {author} {\bibfnamefont {J.}~\bibnamefont
  {Gao}}, \bibinfo {author} {\bibfnamefont {M.}~\bibnamefont {Daal}}, \bibinfo
  {author} {\bibfnamefont {J.~M.}\ \bibnamefont {Martinis}}, \bibinfo {author}
  {\bibfnamefont {A.}~\bibnamefont {Vayonakis}}, \bibinfo {author}
  {\bibfnamefont {J.}~\bibnamefont {Zmuidzinas}}, \bibinfo {author}
  {\bibfnamefont {B.}~\bibnamefont {Sadoulet}}, \bibinfo {author}
  {\bibfnamefont {B.~A.}\ \bibnamefont {Mazin}}, \bibinfo {author}
  {\bibfnamefont {P.~K.}\ \bibnamefont {Day}}, \ and\ \bibinfo {author}
  {\bibfnamefont {H.~G.}\ \bibnamefont {Leduc}},\ }\bibfield  {title} {\enquote
  {\bibinfo {title} {A semiempirical model for two-level system noise in
  superconducting microresonators},}\ }\href@noop {} {\bibfield  {journal}
  {\bibinfo  {journal} {Appl. Phys. Lett.}\ }\textbf {\bibinfo {volume} {92}},\
  \bibinfo {pages} {212504} (\bibinfo {year} {2008})}\BibitemShut {NoStop}%
\bibitem [{\citenamefont {Wenner}\ \emph {et~al.}(2011)\citenamefont {Wenner},
  \citenamefont {Barends}, \citenamefont {Bialczak}, \citenamefont {Chen},
  \citenamefont {Kelly}, \citenamefont {Lucero}, \citenamefont {Mariantoni},
  \citenamefont {Megrant}, \citenamefont {O'Malley}, \citenamefont {Sank},
  \citenamefont {Vainsencher}, \citenamefont {Wang}, \citenamefont {White},
  \citenamefont {Yin}, \citenamefont {Zhao}, \citenamefont {Cleland},\ and\
  \citenamefont {Martinis}}]{Wenner2011}%
  \BibitemOpen
  \bibfield  {author} {\bibinfo {author} {\bibfnamefont {J.}~\bibnamefont
  {Wenner}}, \bibinfo {author} {\bibfnamefont {R.}~\bibnamefont {Barends}},
  \bibinfo {author} {\bibfnamefont {R.~C.}\ \bibnamefont {Bialczak}}, \bibinfo
  {author} {\bibfnamefont {Y.}~\bibnamefont {Chen}}, \bibinfo {author}
  {\bibfnamefont {J.}~\bibnamefont {Kelly}}, \bibinfo {author} {\bibfnamefont
  {E.}~\bibnamefont {Lucero}}, \bibinfo {author} {\bibfnamefont
  {M.}~\bibnamefont {Mariantoni}}, \bibinfo {author} {\bibfnamefont
  {A.}~\bibnamefont {Megrant}}, \bibinfo {author} {\bibfnamefont {P.~J.~J.}\
  \bibnamefont {O'Malley}}, \bibinfo {author} {\bibfnamefont {D.}~\bibnamefont
  {Sank}}, \bibinfo {author} {\bibfnamefont {A.}~\bibnamefont {Vainsencher}},
  \bibinfo {author} {\bibfnamefont {H.}~\bibnamefont {Wang}}, \bibinfo {author}
  {\bibfnamefont {T.~C.}\ \bibnamefont {White}}, \bibinfo {author}
  {\bibfnamefont {Y.}~\bibnamefont {Yin}}, \bibinfo {author} {\bibfnamefont
  {J.}~\bibnamefont {Zhao}}, \bibinfo {author} {\bibfnamefont {A.~N.}\
  \bibnamefont {Cleland}}, \ and\ \bibinfo {author} {\bibfnamefont {J.~M.}\
  \bibnamefont {Martinis}},\ }\bibfield  {title} {\enquote {\bibinfo {title}
  {Surface loss simulations of superconducting coplanar waveguide
  resonators},}\ }\href {\doibase http://dx.doi.org/10.1063/1.3637047}
  {\bibfield  {journal} {\bibinfo  {journal} {Appl. Phys. Lett.}\ }\textbf
  {\bibinfo {volume} {99}},\ \bibinfo {eid} {113513} (\bibinfo {year}
  {2011})}\BibitemShut {NoStop}%
\bibitem [{\citenamefont {Neto}(2010)}]{Neto2010a}%
  \BibitemOpen
  \bibfield  {author} {\bibinfo {author} {\bibfnamefont {A.}~\bibnamefont
  {Neto}},\ }\bibfield  {title} {\enquote {\bibinfo {title} {Uwb, non
  dispersive radiation from the planarly fed leaky lens antenna; part 1: Theory
  and design},}\ }\href {\doibase 10.1109/TAP.2010.2048879} {\bibfield
  {journal} {\bibinfo  {journal} {IEEE Trans. Antennas Propag.}\ }\textbf
  {\bibinfo {volume} {58}},\ \bibinfo {pages} {2238--2247} (\bibinfo {year}
  {2010})}\BibitemShut {NoStop}%
\bibitem [{\citenamefont {{Yurduseven}}, \citenamefont {{Llombart Juan}},\ and\
  \citenamefont {{Neto}}(2016)}]{Ozan2016}%
  \BibitemOpen
  \bibfield  {author} {\bibinfo {author} {\bibfnamefont {O.}~\bibnamefont
  {{Yurduseven}}}, \bibinfo {author} {\bibfnamefont {N.}~\bibnamefont
  {{Llombart Juan}}}, \ and\ \bibinfo {author} {\bibfnamefont {A.}~\bibnamefont
  {{Neto}}},\ }\bibfield  {title} {\enquote {\bibinfo {title} {{A
  Dual-Polarized Leaky Lens Antenna for Wideband Focal Plane Arrays}},}\ }\href
  {\doibase 10.1109/TAP.2016.2574903} {\bibfield  {journal} {\bibinfo
  {journal} {IEEE Transactions on Antennas and Propagation}\ }\textbf {\bibinfo
  {volume} {64}},\ \bibinfo {pages} {3330--3337} (\bibinfo {year}
  {2016})}\BibitemShut {NoStop}%
\bibitem [{\citenamefont {{Mirzaei}}\ \emph {et~al.}(2021)\citenamefont
  {{Mirzaei}}, \citenamefont {{Barrentine}}, \citenamefont {{Bulcha}},
  \citenamefont {{Cataldo}}, \citenamefont {{Connors}}, \citenamefont
  {{Ehsan}}, \citenamefont {{Essinger-Hileman}}, \citenamefont {{Hess}},
  \citenamefont {{Mugge-Durum}}, \citenamefont {{Noroozian}}, \citenamefont
  {{Oxholm}}, \citenamefont {{Stevenson}}, \citenamefont {{Switzer}},
  \citenamefont {{Volpert}},\ and\ \citenamefont {{Wollack}}}]{Mirzaei2021}%
  \BibitemOpen
  \bibfield  {author} {\bibinfo {author} {\bibfnamefont {M.}~\bibnamefont
  {{Mirzaei}}}, \bibinfo {author} {\bibfnamefont {E.~M.}\ \bibnamefont
  {{Barrentine}}}, \bibinfo {author} {\bibfnamefont {B.~T.}\ \bibnamefont
  {{Bulcha}}}, \bibinfo {author} {\bibfnamefont {G.}~\bibnamefont {{Cataldo}}},
  \bibinfo {author} {\bibfnamefont {J.~A.}\ \bibnamefont {{Connors}}}, \bibinfo
  {author} {\bibfnamefont {N.}~\bibnamefont {{Ehsan}}}, \bibinfo {author}
  {\bibfnamefont {T.~M.}\ \bibnamefont {{Essinger-Hileman}}}, \bibinfo {author}
  {\bibfnamefont {L.~A.}\ \bibnamefont {{Hess}}}, \bibinfo {author}
  {\bibfnamefont {J.~W.}\ \bibnamefont {{Mugge-Durum}}}, \bibinfo {author}
  {\bibfnamefont {O.}~\bibnamefont {{Noroozian}}}, \bibinfo {author}
  {\bibfnamefont {T.~M.}\ \bibnamefont {{Oxholm}}}, \bibinfo {author}
  {\bibfnamefont {T.~R.}\ \bibnamefont {{Stevenson}}}, \bibinfo {author}
  {\bibfnamefont {E.~R.}\ \bibnamefont {{Switzer}}}, \bibinfo {author}
  {\bibfnamefont {C.~G.}\ \bibnamefont {{Volpert}}}, \ and\ \bibinfo {author}
  {\bibfnamefont {E.~J.}\ \bibnamefont {{Wollack}}},\ }\bibfield  {title}
  {\enquote {\bibinfo {title} {{Mu-Spec Spectrometers for the EXCLAIM
  Instrument}},}\ }\href@noop {} {\bibfield  {journal} {\bibinfo  {journal}
  {arXiv e-prints}\ ,\ \bibinfo {eid} {arXiv:2101.11741}} (\bibinfo {year}
  {2021})},\ \Eprint {http://arxiv.org/abs/2101.11741} {arXiv:2101.11741
  [astro-ph.IM]} \BibitemShut {NoStop}%
\bibitem [{\citenamefont {{Thompson}}(2012)}]{Thompson2012}%
  \BibitemOpen
  \bibfield  {author} {\bibinfo {author} {\bibfnamefont {C.~V.}\ \bibnamefont
  {{Thompson}}},\ }\bibfield  {title} {\enquote {\bibinfo {title} {{Solid-State
  Dewetting of Thin Films}},}\ }\href {\doibase
  10.1146/annurev-matsci-070511-155048} {\bibfield  {journal} {\bibinfo
  {journal} {Annual Review of Materials Research}\ }\textbf {\bibinfo {volume}
  {42}},\ \bibinfo {pages} {399--434} (\bibinfo {year} {2012})}\BibitemShut
  {NoStop}%
\bibitem [{\citenamefont {Zhang}\ \emph {et~al.}(2019)\citenamefont {Zhang},
  \citenamefont {Dabironezare}, \citenamefont {Carluccio}, \citenamefont
  {Neto},\ and\ \citenamefont {Llombart}}]{Huasheng2019}%
  \BibitemOpen
  \bibfield  {author} {\bibinfo {author} {\bibfnamefont {H.}~\bibnamefont
  {Zhang}}, \bibinfo {author} {\bibfnamefont {S.~O.}\ \bibnamefont
  {Dabironezare}}, \bibinfo {author} {\bibfnamefont {G.}~\bibnamefont
  {Carluccio}}, \bibinfo {author} {\bibfnamefont {A.}~\bibnamefont {Neto}}, \
  and\ \bibinfo {author} {\bibfnamefont {N.}~\bibnamefont {Llombart}},\
  }\bibfield  {title} {\enquote {\bibinfo {title} {A go/fo tool for analyzing
  quasi-optical systems in reception},}\ }in\ \href {\doibase
  10.1109/IRMMW-THz.2019.8873950} {\emph {\bibinfo {booktitle} {2019 44th
  International Conference on Infrared, Millimeter, and Terahertz Waves
  (IRMMW-THz)}}}\ (\bibinfo {year} {2019})\ pp.\ \bibinfo {pages}
  {1--2}\BibitemShut {NoStop}%
\bibitem [{\citenamefont {{Ferrari}}\ \emph {et~al.}(2018)\citenamefont
  {{Ferrari}}, \citenamefont {{Yates}}, \citenamefont {{Eggens}}, \citenamefont
  {{Baryshev}},\ and\ \citenamefont {{Baselmans}}}]{Ferrari2018}%
  \BibitemOpen
  \bibfield  {author} {\bibinfo {author} {\bibfnamefont {L.}~\bibnamefont
  {{Ferrari}}}, \bibinfo {author} {\bibfnamefont {S.~J.~C.}\ \bibnamefont
  {{Yates}}}, \bibinfo {author} {\bibfnamefont {M.}~\bibnamefont {{Eggens}}},
  \bibinfo {author} {\bibfnamefont {A.~M.}\ \bibnamefont {{Baryshev}}}, \ and\
  \bibinfo {author} {\bibfnamefont {J.~J.~A.}\ \bibnamefont {{Baselmans}}},\
  }\bibfield  {title} {\enquote {\bibinfo {title} {{MKID Large Format Array
  Testbed}},}\ }\href {\doibase 10.1109/TTHZ.2018.2871365} {\bibfield
  {journal} {\bibinfo  {journal} {IEEE Transactions on Terahertz Science and
  Technology}\ }\textbf {\bibinfo {volume} {8}},\ \bibinfo {pages} {572--580}
  (\bibinfo {year} {2018})}\BibitemShut {NoStop}%
\bibitem [{\citenamefont {de~Visser}\ \emph {et~al.}(2012)\citenamefont
  {de~Visser}, \citenamefont {Baselmans}, \citenamefont {Yates}, \citenamefont
  {Diener}, \citenamefont {Endo},\ and\ \citenamefont
  {Klapwijk}}]{Visser2012b}%
  \BibitemOpen
  \bibfield  {author} {\bibinfo {author} {\bibfnamefont {P.~J.}\ \bibnamefont
  {de~Visser}}, \bibinfo {author} {\bibfnamefont {J.~J.~A.}\ \bibnamefont
  {Baselmans}}, \bibinfo {author} {\bibfnamefont {S.~J.~C.}\ \bibnamefont
  {Yates}}, \bibinfo {author} {\bibfnamefont {P.}~\bibnamefont {Diener}},
  \bibinfo {author} {\bibfnamefont {A.}~\bibnamefont {Endo}}, \ and\ \bibinfo
  {author} {\bibfnamefont {T.~M.}\ \bibnamefont {Klapwijk}},\ }\bibfield
  {title} {\enquote {\bibinfo {title} {Microwave-induced excess quasiparticles
  in superconducting resonators measured through correlated conductivity
  fluctuations},}\ }\href@noop {} {\bibfield  {journal} {\bibinfo  {journal}
  {Appl. Phys. Lett.}\ }\textbf {\bibinfo {volume} {100}},\ \bibinfo {pages}
  {162601} (\bibinfo {year} {2012})}\BibitemShut {NoStop}%
\bibitem [{\citenamefont {Janssen}\ \emph {et~al.}(2013)\citenamefont
  {Janssen}, \citenamefont {Baselmans}, \citenamefont {Endo}, \citenamefont
  {Ferrari}, \citenamefont {Yates}, \citenamefont {Baryshev},\ and\
  \citenamefont {Klapwijk}}]{Janssen2013}%
  \BibitemOpen
  \bibfield  {author} {\bibinfo {author} {\bibfnamefont {R.~M.~J.}\
  \bibnamefont {Janssen}}, \bibinfo {author} {\bibfnamefont {J.~J.~A.}\
  \bibnamefont {Baselmans}}, \bibinfo {author} {\bibfnamefont {A.}~\bibnamefont
  {Endo}}, \bibinfo {author} {\bibfnamefont {L.}~\bibnamefont {Ferrari}},
  \bibinfo {author} {\bibfnamefont {S.~J.~C.}\ \bibnamefont {Yates}}, \bibinfo
  {author} {\bibfnamefont {A.~M.}\ \bibnamefont {Baryshev}}, \ and\ \bibinfo
  {author} {\bibfnamefont {T.~M.}\ \bibnamefont {Klapwijk}},\ }\bibfield
  {title} {\enquote {\bibinfo {title} {High optical efficiency and photon noise
  limited sensitivity of microwave kinetic inductance detectors using phase
  readout},}\ }\href {\doibase http://dx.doi.org/10.1063/1.4829657} {\bibfield
  {journal} {\bibinfo  {journal} {Appl. Phys. Lett.}\ }\textbf {\bibinfo
  {volume} {103}},\ \bibinfo {eid} {203503} (\bibinfo {year}
  {2013})}\BibitemShut {NoStop}%
\bibitem [{\citenamefont {Baselmans}\ \emph {et~al.}(2008)\citenamefont
  {Baselmans}, \citenamefont {Yates}, \citenamefont {Barends}, \citenamefont
  {Lankwarden}, \citenamefont {Gao}, \citenamefont {Hoevers},\ and\
  \citenamefont {Klapwijk}}]{Baselmans2008}%
  \BibitemOpen
  \bibfield  {author} {\bibinfo {author} {\bibfnamefont {J.}~\bibnamefont
  {Baselmans}}, \bibinfo {author} {\bibfnamefont {S.~J.~C.}\ \bibnamefont
  {Yates}}, \bibinfo {author} {\bibfnamefont {R.}~\bibnamefont {Barends}},
  \bibinfo {author} {\bibfnamefont {Y.~J.~Y.}\ \bibnamefont {Lankwarden}},
  \bibinfo {author} {\bibfnamefont {J.~R.}\ \bibnamefont {Gao}}, \bibinfo
  {author} {\bibfnamefont {H.}~\bibnamefont {Hoevers}}, \ and\ \bibinfo
  {author} {\bibfnamefont {T.~M.}\ \bibnamefont {Klapwijk}},\ }\bibfield
  {title} {\enquote {\bibinfo {title} {Noise and sensitivity of aluminum
  kinetic inductance detectors for sub-mm astronomy},}\ }\href@noop {}
  {\bibfield  {journal} {\bibinfo  {journal} {J. Low Temp. Phys.}\ }\textbf
  {\bibinfo {volume} {151}},\ \bibinfo {pages} {524--529} (\bibinfo {year}
  {2008})}\BibitemShut {NoStop}%
\bibitem [{\citenamefont {Janssen}\ \emph {et~al.}(2014)\citenamefont
  {Janssen}, \citenamefont {Endo}, \citenamefont {de~Visser}, \citenamefont
  {Klapwijk},\ and\ \citenamefont {Baselmans}}]{Janssen2014b}%
  \BibitemOpen
  \bibfield  {author} {\bibinfo {author} {\bibfnamefont {R.~M.~J.}\
  \bibnamefont {Janssen}}, \bibinfo {author} {\bibfnamefont {A.}~\bibnamefont
  {Endo}}, \bibinfo {author} {\bibfnamefont {P.~J.}\ \bibnamefont {de~Visser}},
  \bibinfo {author} {\bibfnamefont {T.~M.}\ \bibnamefont {Klapwijk}}, \ and\
  \bibinfo {author} {\bibfnamefont {J.~J.~A.}\ \bibnamefont {Baselmans}},\
  }\bibfield  {title} {\enquote {\bibinfo {title} {Equivalence of optical and
  electrical noise equivalent power of hybrid {NbTiN-Al} microwave kinetic
  inductance detectors},}\ }\href {\doibase
  http://dx.doi.org/10.1063/1.4901733} {\bibfield  {journal} {\bibinfo
  {journal} {Appl. Phys. Lett.}\ }\textbf {\bibinfo {volume} {105}},\ \bibinfo
  {eid} {193504} (\bibinfo {year} {2014})}\BibitemShut {NoStop}%
\bibitem [{\citenamefont {{de Rooij}}\ \emph {et~al.}(2021)\citenamefont {{de
  Rooij}}, \citenamefont {{Baselmans}}, \citenamefont {{Murugesan}},
  \citenamefont {{Thoen}},\ and\ \citenamefont {{de Visser}}}]{2021deRooij}%
  \BibitemOpen
  \bibfield  {author} {\bibinfo {author} {\bibfnamefont {S.~A.~H.}\
  \bibnamefont {{de Rooij}}}, \bibinfo {author} {\bibfnamefont {J.~J.~A.}\
  \bibnamefont {{Baselmans}}}, \bibinfo {author} {\bibfnamefont
  {V.}~\bibnamefont {{Murugesan}}}, \bibinfo {author} {\bibfnamefont {D.~J.}\
  \bibnamefont {{Thoen}}}, \ and\ \bibinfo {author} {\bibfnamefont {P.~J.}\
  \bibnamefont {{de Visser}}},\ }\bibfield  {title} {\enquote {\bibinfo {title}
  {{Strong reduction of quasiparticle fluctuations in a superconductor due to
  decoupling of the quasiparticle number and lifetime}},}\ }\href {\doibase
  10.1103/PhysRevB.104.L180506} {\bibfield  {journal} {\bibinfo  {journal}
  {Physical Review B}\ }\textbf {\bibinfo {volume} {104}},\ \bibinfo {eid}
  {L180506} (\bibinfo {year} {2021})},\ \Eprint
  {http://arxiv.org/abs/2103.04777} {arXiv:2103.04777 [cond-mat.supr-con]}
  \BibitemShut {NoStop}%
\bibitem [{\citenamefont {{Moseley}}, \citenamefont {{Mather}},\ and\
  \citenamefont {{McCammon}}(1984)}]{Moseley1984}%
  \BibitemOpen
  \bibfield  {author} {\bibinfo {author} {\bibfnamefont {S.~H.}\ \bibnamefont
  {{Moseley}}}, \bibinfo {author} {\bibfnamefont {J.~C.}\ \bibnamefont
  {{Mather}}}, \ and\ \bibinfo {author} {\bibfnamefont {D.}~\bibnamefont
  {{McCammon}}},\ }\bibfield  {title} {\enquote {\bibinfo {title} {{Thermal
  detectors as x-ray spectrometers}},}\ }\href {\doibase 10.1063/1.334129}
  {\bibfield  {journal} {\bibinfo  {journal} {Journal of Applied Physics}\
  }\textbf {\bibinfo {volume} {56}},\ \bibinfo {pages} {1257--1262} (\bibinfo
  {year} {1984})}\BibitemShut {NoStop}%
\bibitem [{\citenamefont {Baselmans}\ \emph {et~al.}(2022)\citenamefont
  {Baselmans}, \citenamefont {Facchin}, \citenamefont {Pascual~Laguna},
  \citenamefont {Bueno}, \citenamefont {Thoen}, \citenamefont {Murugesan},
  \citenamefont {Llombart},\ and\ \citenamefont {de~Visser}}]{2022Zenodo}%
  \BibitemOpen
  \bibfield  {author} {\bibinfo {author} {\bibfnamefont {J.}~\bibnamefont
  {Baselmans}}, \bibinfo {author} {\bibfnamefont {F.}~\bibnamefont {Facchin}},
  \bibinfo {author} {\bibfnamefont {A.}~\bibnamefont {Pascual~Laguna}},
  \bibinfo {author} {\bibfnamefont {J.}~\bibnamefont {Bueno}}, \bibinfo
  {author} {\bibfnamefont {D.}~\bibnamefont {Thoen}}, \bibinfo {author}
  {\bibfnamefont {V.}~\bibnamefont {Murugesan}}, \bibinfo {author}
  {\bibfnamefont {N.}~\bibnamefont {Llombart}}, \ and\ \bibinfo {author}
  {\bibfnamefont {P.}~\bibnamefont {de~Visser}},\ }\bibfield  {title} {\enquote
  {\bibinfo {title} {Data for: "ultra-sensitive thz microwave kinetic
  inductance detectors for future space telescopes"},}\ }\href@noop {}
  {\bibfield  {journal} {\bibinfo  {journal}
  {https://doi.org/10.5281/zenodo.6607050}\ } (\bibinfo {year}
  {2022})}\BibitemShut {NoStop}%
\end{thebibliography}%
\clearpage
\onecolumngrid

\section{Appendix}

\subsection{Cross power spectral density and lifetime fits}\label{Appendix1}
\begin{figure}[!h]
\centering
\includegraphics[width=0.5\linewidth]{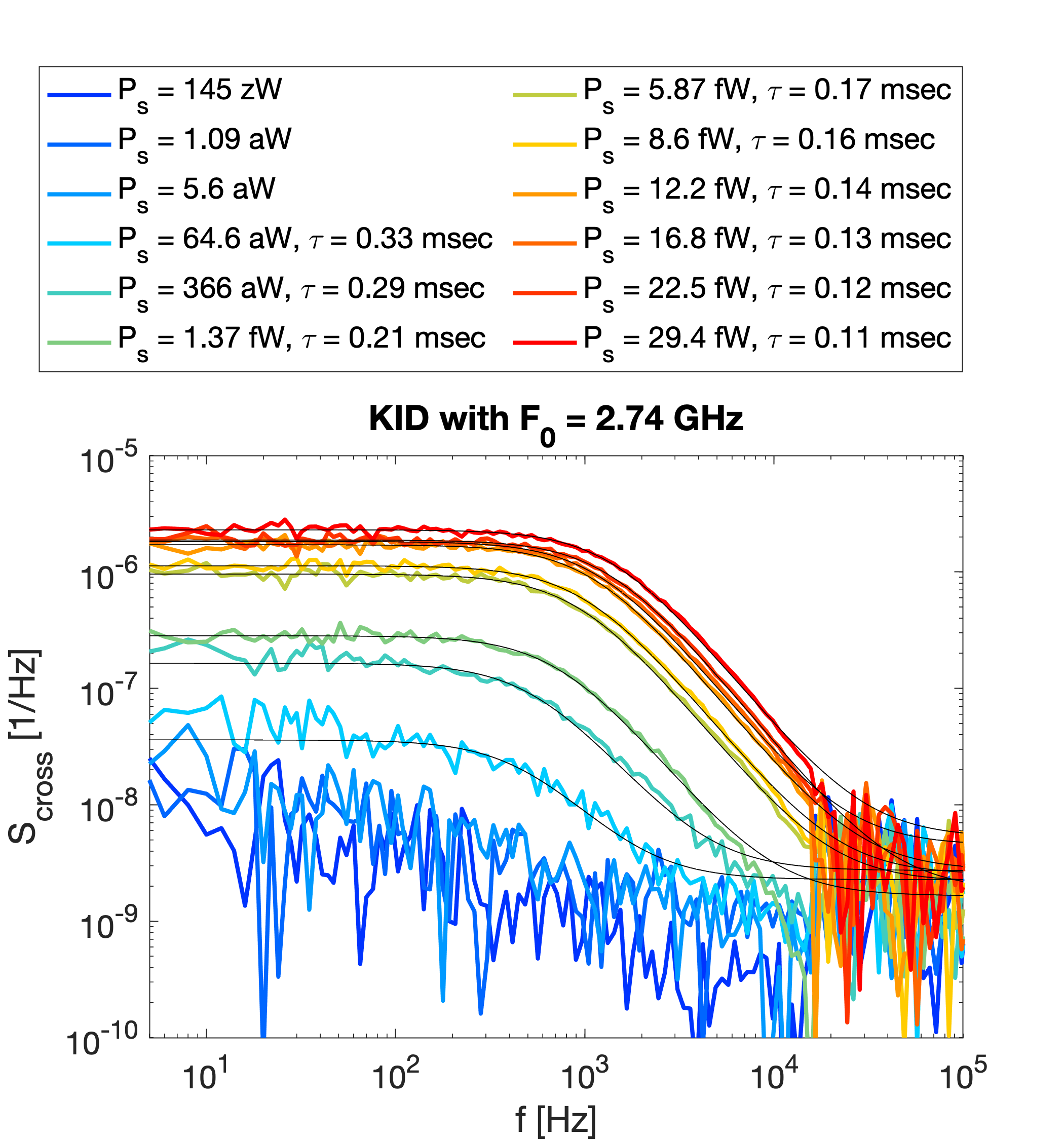}
\caption{\label{Fig:A1} Cross power spectral density at each source power and the fits resulting in the apparent recombination time as indicated. Only at the lowest radiator powers does the photon-noise signal disappear and we cannot fit the data with a single Lorentzian roll-off. }
\end{figure}

In Fig. \ref{Fig:3}a of the main text we show in the insert the apparent quasiparticle recombination time $\tau^*_R$. These datapoints were obtained from the Lorentzian fits to the cross power spectral density of the noise measured at each radiator power. The data and fits that resulted in the values of the recombination time are given in Fig.\ref{Fig:A1}.
\clearpage
\subsection{Energy resolution}\label{Appendix2}

\begin{figure}[!h]
\centering
\includegraphics[width=0.5\linewidth]{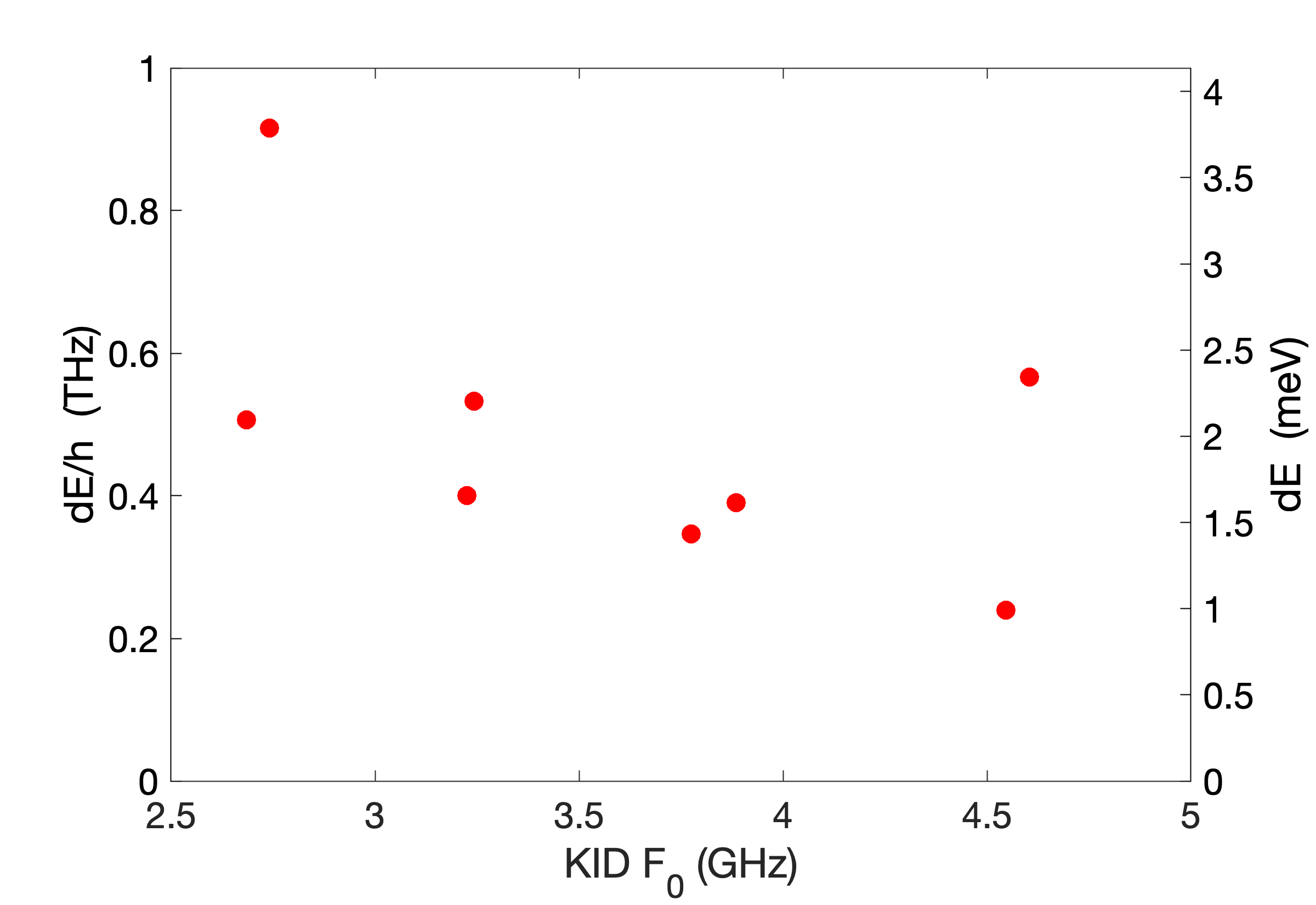}
\caption{\label{Fig:A3}  Estimated energy resolution for all MKIDs discussed in this work, given as a function of the MKID resonance frequency.}
\end{figure}

Using NEP$_{exp}(P_{abs},f\:=\:\mathrm{200\:Hz})$ and the lifetimes obtained from the fits to the photon noise cross power spectral density as discussed above, we can obtain an estimate of the energy resolution, $dE=NEP_{exp}(P_{abs},f\:=\:\mathrm{200\:Hz})\cdot\sqrt{\tau^*_R}$. The result is given in Fig. \ref{Fig:A3}. We find an energy resolution averaged over all MKIDs of $dE/h $ = 0.5 $\pm$ 0.2 THz and a clear improvement towards the smaller device volumes, which results from the fact that the smaller devices have a shorter quasiparticle lifetime and a similar NEP. 

\subsection{NEP$_{exp}(P_{abs},f\:=\:\mathrm{200\:Hz})$ as function of absorbed power}\label{Appendix3}
\begin{figure*}[!ht]
\centering
\includegraphics[width=1\textwidth]{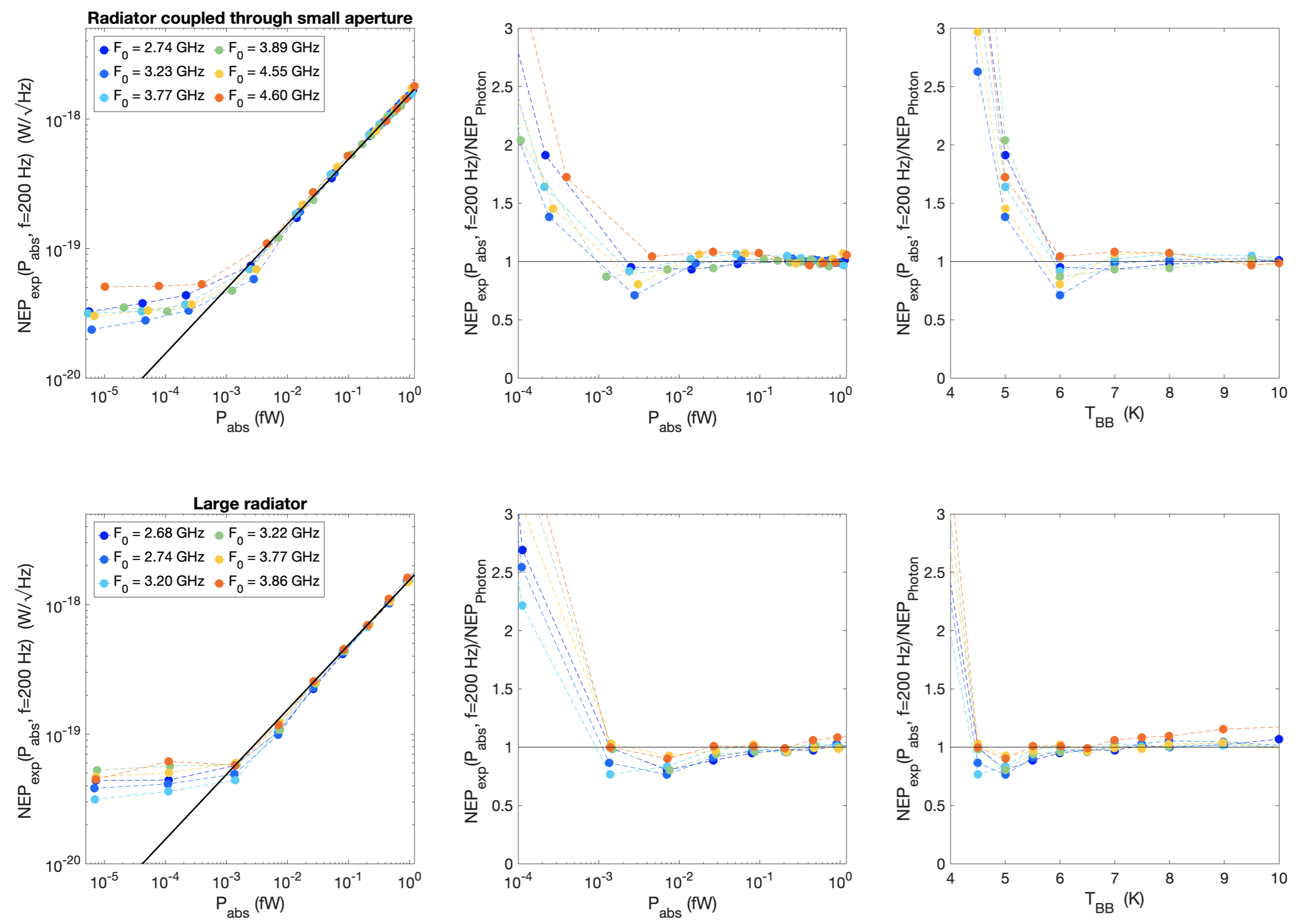}
\caption{\label{Fig:A2} NEP as function of absorbed power for all measured devices. Top row: Data of all measured MKIDs using radiation coupling via a small pinhole to the small radiator source. The left figure gives the data for all KIDs identical to Fig3.e in the main text. The centre and right plots give the NEP divided by the photon noise limit, showing more clearly that the measured NEP can be below the photon noise limit for 10 aW $<\:P_{abs}\:<$ 100 aW. Bottom row: Similar data, but now using the large radiator coupled without pinhole to the detector chip. We observe that the region where the experimental NEP is below the photon noise limit scales with $P_{abs}$, and not with radiator temperature.}
\end{figure*}
In Fig. \ref{Fig:4}a of the main manuscript we give NEP$_{exp}(P_{abs},f\:=\:\mathrm{200\:Hz})$ for the MKID with a resonant frequency of 2.74 GHz. The other measured MKIDs show a similar behaviour, as shown in the top left figure of Fig. \ref{Fig:A2}. However, in the power interval between 10 and 100 aW, we see for most detectors that the measured NEP$_{exp}(P_{abs},f\:=\:\mathrm{200\:Hz})$ is slightly lower than the photon noise limit. We re-plot the same data normalised to the photon noise limit in the other two panels in the top row to make this more clear. To experimentally address this issue, we repeat the measurement using a large (40 mm diameter)\ radiator coupled to the detector chip without the small aperture used in the rest of the experiments. We observe an identical behaviour, which shows both the same limited NEP and an NEP lower than the photon noise limit in the same 10-100 aW absorbed power range. This correspondence in power and not in temperature very clearly excludes possible issues with the filter stack or thermometer: The coupling efficiency with the large black body is 1.27, whereas it is 7\% with the aperture, that is, about a factor 16 difference This implies that the same absorbed power levels are obtained at different radiator temperatures. This is very clear from the rightmost figures: At 6K we observe in the top panel that NEP$_{exp}(P_{abs},f\:=\:\mathrm{200\:Hz})$ is below the photon noise limit. However, in the bottom panel NEP$_{exp}(P_{abs},f\:=\:\mathrm{200\:Hz})$ is identical to the background limit. To further exclude issues with the thermometer, we recalibrated our thermometer as discussed in the main text. The exact mechanism of the discussed effect remains uncertain, but it could be related to the fact that the photon arrival rate is  1 kHz for 1 aW, respectively. This is of the order of the bandwidth given by the recombination time, $\frac{1}{2\pi\tau^*_R}$, which is 500 Hz for $\tau^*_R$=0.3 msec. 

\end{document}